\documentclass[fleqn,usenatbib]{mnras}

\usepackage{newtxtext,newtxmath}
\usepackage{comment}
\usepackage{listings}
\usepackage[T1]{fontenc}
\usepackage{xcolor, color}
\usepackage{lineno}
\DeclareRobustCommand{\VAN}[3]{#2}
\let\VANthebibliography\thebibliography
\def\thebibliography{\DeclareRobustCommand{\VAN}[3]{##3}\VANthebibliography}

\usepackage{graphicx}	% Including figure files
\usepackage{amsmath}	% Advanced maths commands
\usepackage{bm}
\usepackage{physics}
\usepackage{parskip}% http://ctan.org/pkg/parskip
\newcommand{\tempoDOS}{$\mathrm{{\scriptstyle TEMPO2}}$}

\newcommand{\temponest}{$\mathrm{{\scriptstyle TEMPONEST}}$}
\title[Apparent dispersion in pulsar braking indices]{Apparent dispersion in pulsar braking index measurements caused by timing noise}

\author[A.~F.~Vargas et al.]{\parbox{\linewidth}{\centering
Andr\'es F. Vargas$^{1,2}$\thanks{E-mail: afvargas@student.unimelb.edu.au}, and
Andrew Melatos$^{1,2}$
}\\\\
$^{1}$School of Physics, University of Melbourne, Parkville, VIC 3010, Australia\\
$^{2}$OzGrav: The ARC Centre of Excellence for Gravitational-wave Discovery, University of Melbourne, Parkville, VIC 3010, Australia
}

\date{Accepted XXX. Received YYY; in original form ZZZ}

\pubyear{2022}

\begin{document}
\label{firstpage}
\pagerange{\pageref{firstpage}--\pageref{lastpage}}
\maketitle

\begin{abstract}

\noindent Stochastic temporal wandering of the spin frequency $\nu$ of a rotation-powered pulsar (i.e.~the achromatic component of timing noise unrelated to interstellar propagation) affects the accuracy with which the secular braking torque can be measured. Observational studies confirm that pulsars with anomalous braking indices $\vert n \vert = \vert \nu \ddot{\nu} / \dot{\nu}^2 \vert \gg 1$ exhibit elevated levels of timing noise, where an overdot symbolizes a derivative with respect to time. Here it is shown, through analytic calculations and Monte Carlo simulations involving synthetic data and modern Bayesian timing techniques, that the variance $\langle n^2 \rangle$ of the measured $n$ scales with the square of the timing noise amplitude $\sigma_{\ddot{\nu}}$. The anomalous regime $\langle n^2 \rangle \gg 1$ corresponds to $ \sigma_{\ddot{\nu}}^2 \gg 10^{-60} (\gamma_{\ddot{\nu}}/10^{-6} \, {\rm s^{-1}})^2 (\dot{\nu} / 10^{-14} \, {\rm Hz \, s^{-1}})^4 (\nu / 1 \, {\rm Hz})^{-2} (T_{\rm obs} / 10^8 \, {\rm s}) \, {\rm Hz}^2{\rm s}^{-5 }$, where $\gamma_{\ddot{\nu}}$ is a stellar damping time-scale, and $T_{\rm obs}$ is the total observing time. When the inequality in the above condition is reversed, $n$ is dominated by the secular braking torque, and timing measurements return $n\sim 3$, if the secular braking torque is electromagnetic. The variance $\langle n^2 \rangle$ is greater, when the stochastic process driving spin fluctuations differs from the red noise model (e.g.\ power-law spectral density) assumed in the timing solution.\\\\
\end{abstract}

\begin{keywords}
methods: data analysis -- pulsars: general -- stars: rotation
\end{keywords}

%%%%%%%%%%%%%%%%%%%%%%%%%%%%%%%%%%%%%%%%%%%%%%%%%%

%%%%%%%%%%%%%%%%% BODY OF PAPER %%%%%%%%%%%%%%%%%%

\section{Introduction}
\label{Sec:Introduction}

The long-term evolution of the braking torque acting on a rotation-powered pulsar offers insights into the physics of the pulsar's magnetosphere and interior \citep{BlandfordRomani1988}. It can be studied through phase-coherent timing experiments by measuring the braking index,

\begin{equation}
 n = \frac{\nu \ddot{\nu}}{\dot{\nu}^2},
\label{Eq:Intro_n}
\end{equation}

through measuring pulse arrival times, where $\nu(t)$ is the pulse frequency, and an overdot denotes a derivative with respect to time $t$. In the special case where the braking torque is proportional to a power of $\nu$, viz.\ $\dot{\nu} = K \nu^{n_{\rm pl}}$ with $K$ constant, measurements of pulse arrival times yield $n = n_{\rm pl}$ in the absence of stochastic fluctuations in $\nu(t)$ of intrinsic or instrumental origin. Physical examples of power-law braking include vacuum magnetic dipole radiation ($n_{\rm pl}=3$) \citep{GunnOstriker1969}, vacuum electromagnetic radiation for higher-order multipoles with or without general relativistic corrections ($n_{\rm pl} > 3$) \citep{Petri2015,Petri2017}, an extended corotating dipole magnetosphere ($2 \lesssim n_{\rm pl} \leq 3$) \citep{Melatos1997} or its multipolar off-centred generalization \citep{Petri2015,Petri2016}, mass quadrupole gravitational radiation ($n_{\rm pl}=5$) \citep{Thorne1980}, gravitational radiation from r-mode instabilities ($n_{\rm pl}=7$) \citep{PapaloizouPringle1978,Andersson1998,OwenLindblom1998}, and the hydromagnetic torque exerted by an approximately force-free electron-position magnetosphere and relativistic wind ($2 \lesssim n_{\rm pl} \leq 3$) \citep{Goldreich1970,ContopoulosSpitkovsky2006,BucciantiniThompson2006,KouTong2015}. \footnote{It is claimed sometimes that a relativistic wind obeys $n_{\rm pl}=1$. This value applies to the split-monopole geometry \citep{MichelTucker1969}, a theoretical approximation which cannot be realized physically. A realistic wind launched from a dipole magnetosphere obeys $2 \lesssim n_{\rm pl} \leq 3$ \citep{BlandfordRomani1988,Arons1992,BucciantiniThompson2006,ZhangCui2022}.} Phase-coherent timing experiments on pulsars that do not involve major corrections for rotational glitches typically return $2 \lesssim n \leq 3$, consistent with a braking torque of electromagnetic origin \citep{LivingstoneKaspi2007,LivingstoneKaspi2011}, although there are close exceptions such as PSR J1640$-$4631 ($n=3.15\pm0.03$) \citep{ArchibaldGotthelf2016}. An electromagnetic braking torque is also indicated in some glitching pulsars with high-cadence timing data, e.g. PSR J0534$+$2200 and PSR J0835$-$4510, where it is possible to correct for post-glitch recoveries with high accuracy \citep{LynePritchard1996,FuentesEspinoza2017,AkbalGugercinoglu2021}.

A number of rotation-powered pulsars exist, whose measured braking indices are termed `anomalous', because $|n|$ greatly exceeds the standard electromagnetic value, with $3 \ll |n| \lesssim 10^6$ typically, and $n$ is negative in some objects \citep{JohnstonGalloway1999,ChukwudeChidiOdo2016,ParthasarathyJohnston2020,LowerBailes2020}. It is hard to identify a plausible physical mechanism described by $\dot{\nu} = K \nu^{n_{\rm pl}}$ with $K$ constant and $n_{\rm pl} \sim 10^6$. An alternative is that one has $n_{\rm pl} \approx 3$, but $K$ evolves on a time-scale $\tau_{K}$ much shorter than the spin-down time-scale, implying $n = n_{\rm pl} + (\dot{K}/K)(\nu/\dot{\nu}) \approx \nu/(\dot{\nu} \tau_{K}) \gg n_{\rm pl}$. Mechanisms include (counter)alignment of the rotation and magnetic axes \citep{Goldreich1970,LinkEpstein1997,Melatos2000, BarsukovPolyakova2009,JohnstonKarastergiou2017}, magnetic field evolution due to ohmic dissipation or Hall drift \citep{TaurisKonar2001,GeppertRheinhardt2002,PonsVigano2012,GourgouliatosCumming2015}, and precession \citep{BarsukovTsygan2010,BiryukovBeskin2012,GoglichidzeBarsukov2015,WassermanCordes2022}. Another alternative --- which is the focus of this paper --- is that the deterministic, power-law torque is masked by a stochastic process, which dominates $\ddot{\nu}$ over typical observational time-scales. Examples include relaxation processes mediated by crust-superfluid coupling between rotational glitches \citep{AlparBaykal2006,GugercinogluAlpar2014,Akbal2017,Gugercinoglu2017,LowerJohnston2021} and red timing noise intrinsic to the rotation of the stellar crust or superfluid core \citep{CordesDowns1985,AlparNandkumar1986,Jones1990,D'AlessandroMcCulloch1995,MelatosLink2014,ChukwudeChidiOdo2016}, as distinct from timing noise produced by propagation effects in the magnetosphere and interstellar medium \citep{GoncharovReardon2021}. The stochastic process may mask the deterministic braking physically, by adding a fluctuating component to the torque, or observationally, by confounding the measurement of the long-term temporal average of $\ddot{\nu}$, e.g. by contaminating the Taylor coefficients of a polynomial ephemeris \citep{ChukwudeBaiden2010,ColesHobbs2011}. Several population studies confirm that anomalous braking indices are correlated with glitch activity and timing noise amplitude \citep{ArzoumanianNice1994,JohnstonGalloway1999,UramaLink2006,LowerJohnston2021}.

In this paper, we quantify further the role played by stochastic timing noise in masking an underlying, secular braking torque. We run a set of controlled, systematic, Monte Carlo simulations, in which pulse times of arrival (TOAs) are generated synthetically for an ensemble of pulsars following $\ddot{\nu}(t)=\ddot{\nu}_{\rm em}(t)+\zeta(t)$, where $\ddot{\nu}_{\rm em}(t)$ obeys power-law braking, i.e. $\dot{\nu}_{\rm em}(t)=K \nu^{n_{\rm pl}}_{\rm em}(t)$ with $n_{\rm pl} = 3$ (say) and $K$ constant, and $\zeta(t)$ is a phenomenological fluctuating driver, whose power spectral density (PSD) can be white or colored. The measured braking index $n$, defined according to (\ref{Eq:Intro_n}), is estimated from the synthetic data by modern Bayesian methods using the software package \temponest~\citep{ ShannonCordes2010, LentatiAlexander2014,ParthasarathyJohnston2020,LowerJohnston2021}. The goals are to quantify (i) the dispersion in $n$, when $n$ is measured thus in the presence of timing noise, and (ii) the conditions under which one obtains $|n| \gg n_{\rm pl}$. Quantifying (i) and (ii) is the natural next step in extending previous pioneering work on this topic \citep{JohnstonGalloway1999,Chukwude2003,UramaLink2006,ChukwudeBaiden2010,BiryukovBeskin2012,ChukwudeChidiOdo2016,ParthasarathyShannon2019,ParthasarathyJohnston2020,GoncharovReardon2021,LowerJohnston2021}.

The paper is structured as follows. In Section \ref{sec:BI_recovery_scheme} we introduce a simple model to generate noisy time series to emulate TOAs and show how $n$ can be measured with \temponest. In Section \ref{sec:Recovered_n_stats} we conduct an ensemble of systematic numerical experiments with synthetic data to calculate the probability distribution of measured $n$ values as a function of the strength of the timing noise and other properties of its PSD. In Section \ref{sec:Disp_n_vs_noise} and the Appendix, we derive the condition on the noise amplitude that yields $\vert n \vert \gg n_{\rm pl}$ in a population sense. Section \ref{sec:Conclusions} summarizes the conclusions. We emphasize that the results in Sections \ref{sec:Recovered_n_stats} and \ref{sec:Disp_n_vs_noise} cannot be deduced from real data, as there is no way at present to predict theoretically from first principles the statistical properties of the stochastic torque in a real pulsar. The properties can only be inferred indirectly from timing data, and the inference exercise itself is conditional on a model for the stochastic torque, just like in Section~\ref{sec:BI_recovery_scheme} of this paper. However, as a better understanding develops of the underlying physics of timing noise in the future, there is some prospect that the approach in this paper -- with an independent predictive model replacing the one in Section~\ref{sec:BI_recovery_scheme} -- can be applied to real data to disentangle the stochastic and secular contributions to $n$ in individual pulsars.

\section{Simulating braking index measurements}
\label{sec:BI_recovery_scheme}

The twin goals of the paper are to quantify the dispersion in measurements of $n$ caused by timing noise, and to infer a condition on the noise amplitude that yields $\vert n \vert \gg n_{\rm pl}$. To these ends, we perform the following numerical experiment. (i) We create noisy synthetic time series for the dynamical variables in a typical pulsar timing experiment, namely the pulse phase and its first three time derivatives (up to $\ddot{\nu}$), and convert them into TOAs. (ii) We feed the synthetic TOAs into \tempoDOS~\citep{HobbsEdwards2006} and \temponest~\citep{LentatiAlexander2014} to get two independent ``traditional" timing solutions. Two solutions are generated instead of one as a cross-check. (iii) We infer the measured value of $n$ from the synthetic timing solutions and compare it with the injected value of $n_{\rm pl}$. (iv) We repeat steps (i)--(iii) for an ensemble of random realizations of the noise in order to compute the statistics of $n$.

Section~\ref{sec:BI_recovery_scheme} sets out the specific procedures involved in steps (i)--(iv) to assist the interested reader in reproducing the numerical experiment, e.g. to calibrate the interpretation of real data in the future. In Section \ref{subsecII:Noisy_ts_gen} we describe the stochastic model used to generate the synthetic time series. Sections \ref{subsecII:Gen_TOAs_noisy_tn} and \ref{subsecII:Bayesian_param_est_tnest} explain how to convert the synthetic time series into TOAs and track the dynamical variables respectively. Section \ref{subsecII:gen_data_get_n_example} presents an end-to-end worked example of the entire procedure for an arbitrary but representative synthetic pulsar with the rotational parameters of PSR J0942$-$5552.

\subsection{Deterministic and stochastic torques}
\label{subsecII:Noisy_ts_gen}

The electromagnetic pulses emitted by a rotation-powered pulsar are assumed to be phase-locked to the rigid crust and corotating magnetosphere, as long as the emission region does not drift in longitude, i.e. it is assumed in this paper that there is no random walk in the phase $\phi(t)$, beyond what is inherited from the randomly walking frequency via $\phi(t)=\int_{0}^{t}dt'\,\nu(t')$, cf. \cite{CordesHelfand1980}. Hence the spin-down evolution is described in general by four independent variables: the rotational phase $\phi(t)$, the rotational frequency $\nu(t)$ of the crust, and its derivatives $\dot \nu(t)$ and $\ddot \nu(t)$, where an overdot denotes differentiation with respect to time. The dynamical variables are packaged into a column vector ${\bf X} = (X_1,X_2,X_3,X_4)^{\rm T} = (\phi,\nu,\dot{\nu},\ddot{\nu})^{\rm T}$, where ${\rm T}$ denotes the matrix transpose.

The state vector ${\bf X}$ evolves under the action of deterministic and stochastic torques. In this paper, the deterministic torque is taken to be the standard magnetic dipole torque, with $K$ constant and $n_{\rm pl}=3$ \citep{GunnOstriker1969}, but alternatives like a force-free relativistic wind ($n_{\rm pl} \approx 3$) \citep{BucciantiniThompson2006} or mass quadrupole gravitational radiation ($n_{\rm pl}=5$) \citep{Thorne1980} would serve just as well. For simplicity the stochastic torque is taken phenomenologically to be a white noise Langevin term driving $\ddot{\nu}(t)$. Here too there exist valid alternatives, which can be accommodated naturally within the mathematical framework below; the reader is encouraged to experiment, as the need arises. The model also allows for noise to be added to a combination of $\nu(t)$ and/or $\dot{\nu}(t)$, when astronomical data warrant \citep{Cheng1987,Jones1990,MeyersMelatos2021,MeyersO'Neill2021,AntonelliBasu2022}. Importantly, a white noise driver in $\ddot{\nu}(t)$ produces red noise in the measured variables $\phi(t),\nu(t)$ and $\dot{\nu}(t)$, because the white noise is filtered by the damping terms in the Langevin equation. We calibrate the noisy driver to generate phase residuals that resemble qualitatively those observed in real pulsars, as explained in Section~\ref{subsecII:gen_data_get_n_example}. Satisfactory phase residuals can be achieved for drivers described by a range of functional forms.

The above model is described by a set of four simultaneous stochastic differential (Langevin) equations

\begin{equation}
    d\mathbf{X} =  ( \mathbf{A} \, \mathbf{X}+\mathbf{E} )\,dt + \mathbf{\Sigma}\,d \mathbf{B}(t),
    \label{Eq_SecII:model}
\end{equation}

\noindent with 

\begin{equation}
    \bm A = \begin{pmatrix} 0 & 1 & 0 & 0 \\ 0 & -\gamma_{\nu} & 1 & 0\\ 0 & 0 & -\gamma_{\dot{\nu}} & 1 \\
    0 & 0 & 0 & -\gamma_{\ddot{\nu}} \end{pmatrix}, \label{Eq:Amplitudes_Matrix} 
\end{equation}

\begin{equation}
    \bm E = \begin{pmatrix} 0 \\ \gamma_{\nu} \nu_{\rm em}(t) \\ \gamma_{\dot{\nu}} \dot{\nu}_{\rm em}(t) \\ \dddot{\nu}_{\rm em}(t)+\gamma_{\ddot{\nu}} \ddot{\nu}_{\rm em}(t) \end{pmatrix},
\end{equation}

\noindent and

\begin{equation}
    \bm \Sigma = \text{diag}\left(0, 0, 0 ,\sigma_{\Ddot{\nu}}^{2} \right). \label{Eq:Sigma_Matrix}
\end{equation}

In (\ref{Eq_SecII:model})--(\ref{Eq:Sigma_Matrix}), the parameters $\gamma_\nu$, $\gamma_{\dot{\nu}}$, and $\gamma_{\ddot{\nu}}$ are constant damping coefficients, and $\nu_{\rm em}(t)$ is the solution to the deterministic electromagnetic braking law

\begin{equation}
    \dot{\nu}_{\rm em}(t)=K \nu^{n_{\rm pl}}_{\rm em}(t),
    \label{Eq:diff_braking_law}
\end{equation}

with $n_{\rm pl}\approx3$. In~(\ref{Eq_SecII:model}), $d{\bf B}(t)$ denotes the infinitesimal increments of a zero-mean, unit-variance, Wiener process.

It is important to emphasize that (\ref{Eq_SecII:model})--(\ref{Eq:Sigma_Matrix}) are phenomenological equations of motion, in the spirit of previous related work \citep{MeyersMelatos2021,MeyersO'Neill2021,AntonelliBasu2022}. They are not derived from a physical model of timing noise, such as the two-component crust-superfluid model, nor are they unique. Instead, they aim to reproduce qualitatively the observed timing behavior of typical pulsars, namely that $\nu(t) \approx \nu_{\rm em}(t)$ and $\dot{\nu}(t) \approx \dot{\nu}_{\rm em}(t)$ adhere closely to the secular evolution described by (\ref{Eq:diff_braking_law}), with minimal wandering, while $\ddot{\nu}(t)$ wanders randomly and significantly around the secular trend $\langle \ddot{\nu}(t) \rangle = \ddot{\nu}_{\rm em}(t)$. One can see this by writing out the second row of (\ref{Eq_SecII:model}) explicitly as one example:
\begin{equation}
    \frac{d\nu(t)}{dt} = -\gamma_{\nu}[\nu(t)-\nu_{\rm em}(t)]+\dot{\nu}(t).
    \label{Eq:Example_nu_equation}
\end{equation}

It is plain that (\ref{Eq:Example_nu_equation}) describes a relaxation process, in which $\nu(t)$ reverts to the mean $\nu_{\rm em}(t)$ on a time-scale $\gamma_{\nu}^{-1}$. The process is driven by the stochastic torque $\dot{\nu}$, which inherits its randomness from integration of the white-noise driver with amplitude $\sigma_{\ddot{\nu}}$ in (\ref{Eq:Sigma_Matrix}). The fluctuations $| \nu(t) - \nu_{\rm em}(t) |$ are small compared to $\nu_{\rm em}(t)$, provided that $\gamma_{\nu}^{2}\nu_{\rm em}^{2}$ is large compared to ${\rm var}(\dot{\nu})$, the variance of $\dot{\nu}$, a condition which is always observed to hold astrophysically.

The stochastic torques in the components of the Wiener increment $d{\mathbf B}(t)$ are treated as memory-less, white-noise processes with

\begin{equation}
    \langle d\mathbf{B} (t) \rangle = 0,
    \label{Eq_subsecII:whitenoise}
\end{equation}

and 

\begin{equation}
    \langle d \mathbf{B}_{i}(t)d \mathbf{B}_{j}(t') \rangle = \mathbf{\Sigma}_{ij}\delta(t-t')
    \label{Eq_subsecII:memory_less}
\end{equation}

where $\langle ... \rangle$ denotes the ensemble average and $\mathbf{\Sigma}_{ij}$ denotes the $(i,j)$-th element of (\ref{Eq:Sigma_Matrix}). We assume zero cross-correlations for simplicity, with $\mathbf{\Sigma}_{ij}=0$ for $i\neq j$, although it is easy to include off-diagonal elements in ${\bf \Sigma}$ in the future, if warranted by pulsar timing data. We also assume in this paper that the stochastic torque is nonzero for $\ddot{\nu}$ only, with amplitude ${\bf \Sigma}_{4\,4}  = \sigma_{\ddot{\nu}}^{2}$; that is, the other three diagonal components of ${\bf \Sigma}$ vanish. This assumption is made for pragmatic reasons: it ensures that $\nu(t)$, $\dot{\nu}(t)$, and $\ddot{\nu}(t)$ are all differentiable, so that a braking index $n$ can be calculated, and the observationally motivated tests in this paper can be conducted.\footnote{The Wiener increment, $d{\bf B}(t)$, is not differentiable, but its integral $\int_0^t dt' \, d{\bf B}(t')$ is differentiable. As $n$ features derivatives as high as second order, and $n$ is a measurable quantity which must not diverge, one is forced to introduce noise into (\ref{Eq_SecII:model})--(\ref{Eq:Sigma_Matrix}) ``from the bottom up", i.e.\ the white noise driver is added to the right-hand side of $d\ddot{\nu}(t)/dt$ and filtered, through integration, up to $\phi(t)$.} The assumption may not always hold physically; magnetospheric fluctuations could make the pulsar beam and hence $\phi(t)$ wander randomly in longitude, for example (which would imply ${\bf \Sigma}_{1\,1} \neq 0$), over and above any wandering inherited from fluctuations in $\nu(t)$. Henceforth, we refer to the timing noise model described by equations (\ref{Eq_SecII:model})--(\ref{Eq_subsecII:memory_less}) as the Brownian model.

\subsection{Synthetic TOA\lowercase{s}}
\label{subsecII:Gen_TOAs_noisy_tn}

We generate synthetic TOAs from numerical solutions of (\ref{Eq_SecII:model})--(\ref{Eq:diff_braking_law}) by generating a sample of times $t_{i}$, where the phase component $X_1(t_{i})=\phi(t_{i})$ equals a fiducial value (zero without loss of generality). Every numerical solution of (\ref{Eq_SecII:model})--(\ref{Eq:diff_braking_law})  corresponds to one random realization of the system. Solving (\ref{Eq_SecII:model})--(\ref{Eq:diff_braking_law}) again with a new random seed produces a different state sequence ${\bf X}(t_i)$ and hence a different yet statistically equivalent set of TOAs.

To create the TOAs $\{ t_1,\dots, t_{N_{\text{TOA}}}\}$,  we start by selecting randomly a set of $N_{\text{TOA}}$ reference times $\{ t'_1, \dots, t'_{N_{\text{TOA}}} \}$ within the observation interval $0\leq t \leq T_{\text{obs}}$. Random in this context means selected with uniform probability per unit time in the foregoing interval, i.e. according to a Poisson process. We then integrate (\ref{Eq_SecII:model})--(\ref{Eq:diff_braking_law})  numerically from the desired initial conditions $\mathbf{X}(t_{0})$ to obtain $\{\mathbf{X}(t'_{1}),...,\mathbf{X}(t'_{N_{\text{TOA}}})\}$. For each intermediate state $\mathbf{X}(t'_{i})$ we record the time shift  $dt'_{i}$ needed to obtain the next zero crossing of the phase, with $X_{1}(t'_i+dt'_{i})=\phi(t'_{i}+dt'_{i})=0$. The TOAs are then given by $t_{i}=t'_{i}+dt'_{i}$, and the state vector $\mathbf{X}(t'_{i})$ is updated to return $\mathbf{X}(t_{i})$ at the $i$-th TOA. The number of pulses, i.e. the number of times $X_{1}(t)=\phi(t)=0$ occurs in the interval $0 \leq t \leq t_{i} $, is recorded for each TOA $t_{i}$. To finish, all TOAs are reported alongside a constant uncertainty $\Delta_{\rm TOA}$.\footnote{TOAs and their respective measurement uncertainties are reported in neighbouring columns, $t_{i}$ and $\Delta_{{\rm TOA},i}$, in the {\tt .tim} file created by {\tt baboo} (see footnote 4) when generating synthetic data. The uncertainties are processed by~\tempoDOS~and~\temponest~when generating an ephemeris. In this paper, when reporting synthetic TOAs, we set $\Delta_{{\rm TOA},i}=\Delta_{\rm TOA}$ for $1\leq i \leq N_{\rm TOA}$ for simplicity} 

The above procedure is implemented in the publicly available {\tt baboo} package.\footnote{\url{http://www.github.com/meyers-academic/baboo}} The code uses the Runge-Kutta It\^o integrator found in the \texttt{sdeint} python package\footnote{\url{https://github.com/mattja/sdeint}} to solve (\ref{Eq_SecII:model}) numerically~\citep{10.2307/41062628}.

\subsection{Bayesian parameter estimation with \temponest}
\label{subsecII:Bayesian_param_est_tnest}

Every synthetic data realization generated via the procedure in Sections~\ref{subsecII:Noisy_ts_gen} and~\ref{subsecII:Gen_TOAs_noisy_tn} comprises the initial states ${\bf X}(t_{0})$, the pulsar's right ascension (RA) and declination (DEC), and the synthetic TOAs $\{ t_1, \dots, t_{N_{\rm TOA}}\}$ with their uncertainties. These data are fed into \tempoDOS~\citep{HobbsEdwards2006} to generate an initial estimate of the parameters $\theta=\{{\rm RA},{\rm DEC}, \nu,\dot{\nu},\ddot{\nu}\}$ and their respective uncertainties $\Delta \theta = \{ \Delta{\rm RA},\Delta{\rm DEC},\Delta\nu,\Delta\dot{\nu}, \Delta \ddot{\nu}\}$.  

We use the \tempoDOS~output for $\theta$ to set the priors in \temponest. We assume uniform priors for all five components of $\theta$. The prior range for ${\rm RA}$ and ${\rm DEC}$ is $(-10^{-5}~{\rm rad},10^{-5}~{\rm rad})$ around $\mathrm{{\scriptstyle TEMPO2}}$'s central estimates. The prior range for $\nu$ is $(-10^{-4}~{\rm Hz},10^{-4}~{\rm Hz})$ around $\mathrm{{\scriptstyle TEMPO2}}$'s central estimate. For $\dot{\nu}$ we elect to cover the absolute range $(-10^{-12}~{\rm Hz}\,{\rm s}^{-1},0~{\rm Hz}\,{\rm s}^{-1})$, as all the synthetic data generated for this paper refer to hypothetical pulsars with $\dot{\nu} \geq -10^{-12}~{\rm Hz}\,{\rm s}^{-1}$, which are typical of the observed pulsar population~\citep{ManchesterHobbs2005}. For $\ddot{\nu}$ we follow the same approach as in \cite{LowerBailes2020} and \cite{ParthasarathyJohnston2020} and cover the range $(-10^{3} \Delta \ddot{\nu}, 10^{3} \Delta \ddot{\nu})$; the latter range is absolute, it is not centered on~\tempoDOS's central estimate. We follow the same procedure when analyzing an ensemble of random realizations with fixed $\sigma^{2}_{\ddot{\nu}}$, as in Sections~\ref{sec:Recovered_n_stats} and~\ref{sec:Disp_n_vs_noise}, except that the prior on $\ddot{\nu}$ spans the range $(-10^{3} \langle \Delta  \ddot{\nu}\rangle, 10^{3} \langle \Delta \ddot{\nu} \rangle)$, where $\langle \Delta \ddot{\nu} \rangle$ is averaged over the ensemble of realizations. The width of the foregoing priors on all five components of $\theta$ are characteristic of pulsars in the Australia Telescope National Facility (ATNF) pulsar database~\citep{ManchesterHobbs2005}.

\temponest~includes a phenomenological model for the red timing noise in the phase residuals \citep{LentatiAlexander2014}. The phase residuals PSD in the frequency domain is given by
\begin{equation}
    P_{\rm r}(f) = \frac{A^{2}_{\rm red}}{12\pi^{2}}\left(\frac{f}{f_{\rm yr}}\right)^{-\beta}, 
    \label{Eq:Temponest_TimingNoise}
\end{equation}

\noindent where $A_{\rm red}$ is the amplitude, $\beta$ is the spectral index, and we define $f_{\rm yr}=(1~{\rm year})^{-1}$. We remind the reader that equations (\ref{Eq_SecII:model})--(\ref{Eq:Sigma_Matrix}) inject white-noise fluctuations into $\ddot{\nu}$ via $\sigma^{2}_{\ddot{\nu}}$, which are converted into red-noise fluctuations in $\phi,\nu$ and $\dot{\nu}$, when (\ref{Eq_SecII:model})--(\ref{Eq:Sigma_Matrix}) are integrated, through the high-pass filtering action of the damping terms proportional to $\gamma_{\nu}$, $\gamma_{\dot{\nu}}$, and $\gamma_{\ddot{\nu}}$~\citep{MeyersMelatos2021,AntonelliBasu2022}. Excess white noise in the phase residuals, additional to the red noise in~(\ref{Eq:Temponest_TimingNoise}), is handled in~\temponest~by modifying the uncertainties of each TOA according to $\mu=({\rm EQUAD})^{2}+({\rm EFAC})\Delta_{\rm TOA}$, i.e. $\mu$ supersedes $\Delta_{\rm TOA}$. Here, ${\rm EQUAD}$ is the error in quadrature which models stationary excess noise, and ${\rm EFAC}$ is a fitting factor which corrects for unidentified instrumental effects and imperfect estimates of $\Delta_{\rm TOA}$~\citep{LowerBailes2020}. The priors used throughout  the paper for the timing noise ${\rm PSD}$, ${\rm EFAC}$, and ${\rm EQUAD}$ are summarized in Table~\ref{Table:stochastic_params_table}.   

\begin{table}
\caption{Prior ranges for the timing noise parameters used by \temponest.}
\flushleft
\begin{tabular}{p{2cm}p{2.5cm}p{2.5cm}}
\hline
Parameter  [units] & Prior range & Prior type \\
\hline
EFAC & $(-1,3)$ & Uniform \\
EQUAD [s] & $(10^{-10},10^{-2})$ & Log-uniform \\
 $A_{\rm red}~[{\rm yr}^{3/2}]$ & $(10^{-15},10^{-5})$ & Log-uniform \\
$\beta$ & $(2,10)$ & Log-uniform\\
\hline
\end{tabular}
\label{Table:stochastic_params_table}
\end{table}

The \temponest~estimates of $\nu,\dot{\nu}$ and $\ddot{\nu}$ imply a measurement of $n$ via (\ref{Eq:Intro_n}). The uncertainties $\Delta \nu$ and $\Delta \dot{\nu}$ are small compared to $\nu$ and $\dot{\nu}$, e.g. \temponest~yields $\Delta \nu / \nu \sim 10^{-9}$ and $\Delta \dot{\nu}/\dot{\nu} \sim 10^{-3}$ for synthetic data created with $\sigma^{2}_{\ddot{\nu}}=10^{-50}~{\rm Hz}^{2}{\rm s}^{-5}$. Hence the formal uncertainty in the measured $n$, denoted by $\Delta n$, is dominated by the uncertainty in $\ddot{\nu}$, with

\begin{equation}
    \Delta n = \frac{\nu \Delta \ddot{\nu}}{\dot{\nu}^{2}}.
    \label{Eq:Uncertainty_n}
\end{equation}

From a random ensemble of synthetic TOA time series, all created for a given $\sigma_{\ddot{\nu}}$, we construct probability distributions of $n$ and $\Delta n$ following the recipe above. From the distributions, we calculate the average displacement ${\rm ERR}(n)$ of the measured $n$ relative to the injected value of $n_{\rm pl}$. That is, we calculate the average {\em bias} of the measured braking index relative to $n_{\rm pl}$.

Of course, ${\rm ERR}(n)$ cannot be measured astronomically, because observing a real pulsar involves observing a single noise realization (the real one) instead of an ensemble. There is no way to know where the real noise realization falls within the distribution of possible realizations, and so there is no way to know where the actual $n$ measurement falls within the spread of possible $n$ measurements, let alone where it falls relative to $n_{\rm pl}$. In order to quantify the spread of possible $n$ measurements, we calculate the average {\em dispersion} across the ensemble, characterized by the fractional variance

\begin{equation}
    {\rm DISP}(n) = \frac{\langle n^{2} \rangle - n_{\rm pl}^{2}}{n_{\rm pl}^{2}}.
    \label{Eq:Dispersion_n}
\end{equation}

In (\ref{Eq:Dispersion_n}), the average $\langle n^2 \rangle$ is taken over the ensemble of synthetic TOA time series at fixed $\sigma_{\ddot{\nu}}$. ${\rm DISP}(n)$ is a key quantity of interest throughout the rest of this paper,  because it quantifies the fundamental statistical uncertainty associated with measuring $n$ and $n_{\rm pl}$ in a real pulsar, when only one realization (the real one) is available, and we cannot know where it lies in the ensemble distribution. \footnote{${\rm DISP}(n)$ is crudely analogous to the squared standard error of the sample mean in elementary statistics.} It is important to distinguish $\Delta n$ in (\ref{Eq:Uncertainty_n}) and ${\rm DISP}(n)$ in (\ref{Eq:Dispersion_n}). The former quantity is the formal uncertainty returned by~\tempoDOS~or~\temponest, when measuring $n$ from a single noise realization, whereas the latter quantity describes the dispersion of outcomes across the ensemble.  

\subsection{An example: Emulating the representative pulsar PSR J0942$-$5552}
\label{subsecII:gen_data_get_n_example}

\begin{table}
\centering
\caption{Injected rotational parameters and recovered values for the worked example in Section \ref{subsecII:gen_data_get_n_example}.  Values in parentheses indicate the $1\sigma$ uncertainty in the trailing digits. The values of $\nu(t_{0})$ and its derivatives are consistent with $n_{\rm pl}=3$, whereas the recovered values imply $n=5248.85 \neq n_{\rm pl}$. The injected parameters are representative of PSR J0942$-$5552, an arbitrary but typical pulsar in \protect\cite{LowerBailes2020}. The timing analysis assumes the red noise model (\ref{Eq:Temponest_TimingNoise}), while the Brownian model (\ref{Eq_SecII:model})--(\ref{Eq_subsecII:memory_less}) generates the spin fluctuations}. The parameters in the lower half of the table are used to generate the TOA time series.
\label{Table_subsecII:example_injected_values}
\begin{tabular}{llll}
\hline
Parameter & Units & Injected value & Recovered value \\
\hline
 $\nu(t_{0})$ & $\text{Hz}$ & $1.5051430406$ & $1.5051430479(2)$\\
 $\dot{\nu}(t_{0})$ & $10^{-14}~\text{Hz s}^{-1}$ & $-5.1380792001$ & $-5.1(1)$ \\ 
 $\Ddot{\nu}(t_{0})$ & $10^{-24}~\text{Hz s}^{-2}$ & $5.23\times10^{-3}$ & $9(2)$ \\ 
 \hline
 $\gamma_{\nu}$ & ${\rm s}^{-1}$ & $1\times10^{-13}$ & -- \\ 
 $\gamma_{\dot{\nu}}$ & ${\rm s}^{-1}$ & $1\times10^{-13}$ & --\\
 $\gamma_{\ddot{\nu}}$ & ${\rm s}^{-1}$ & $1\times10^{-6}$ & --\\
$\sigma_{\ddot{\nu}}^{2}$ & ${\rm Hz}^{2}{\rm s}^{-5}$ & $2.5\times10^{-50}$  & --\\ 
$T_{\text{obs}}$ & $\text{days}$ & $1.314\times10^{3}$ & -- \\
$N_{\text{TOA}}$ & -- & $1.5\times10^{2}$ & -- \\ 
$\Delta_{\rm TOA}$ & $\mu \text{s}$ & $1\times10^{2}$ & -- \\ 
\hline
\end{tabular}
\end{table}

\begin{figure}
\flushleft
 \includegraphics[width=\columnwidth]{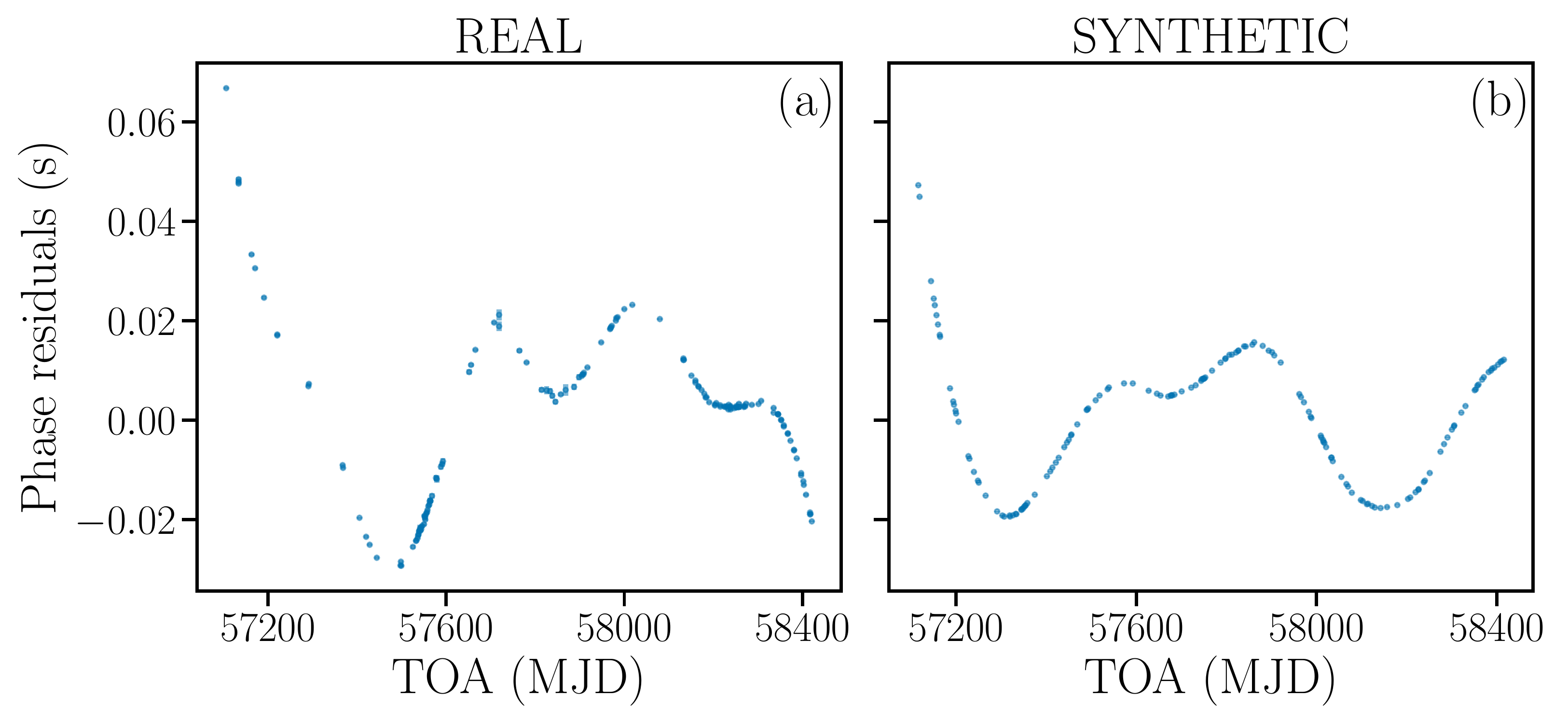}
 \includegraphics[width=\columnwidth]{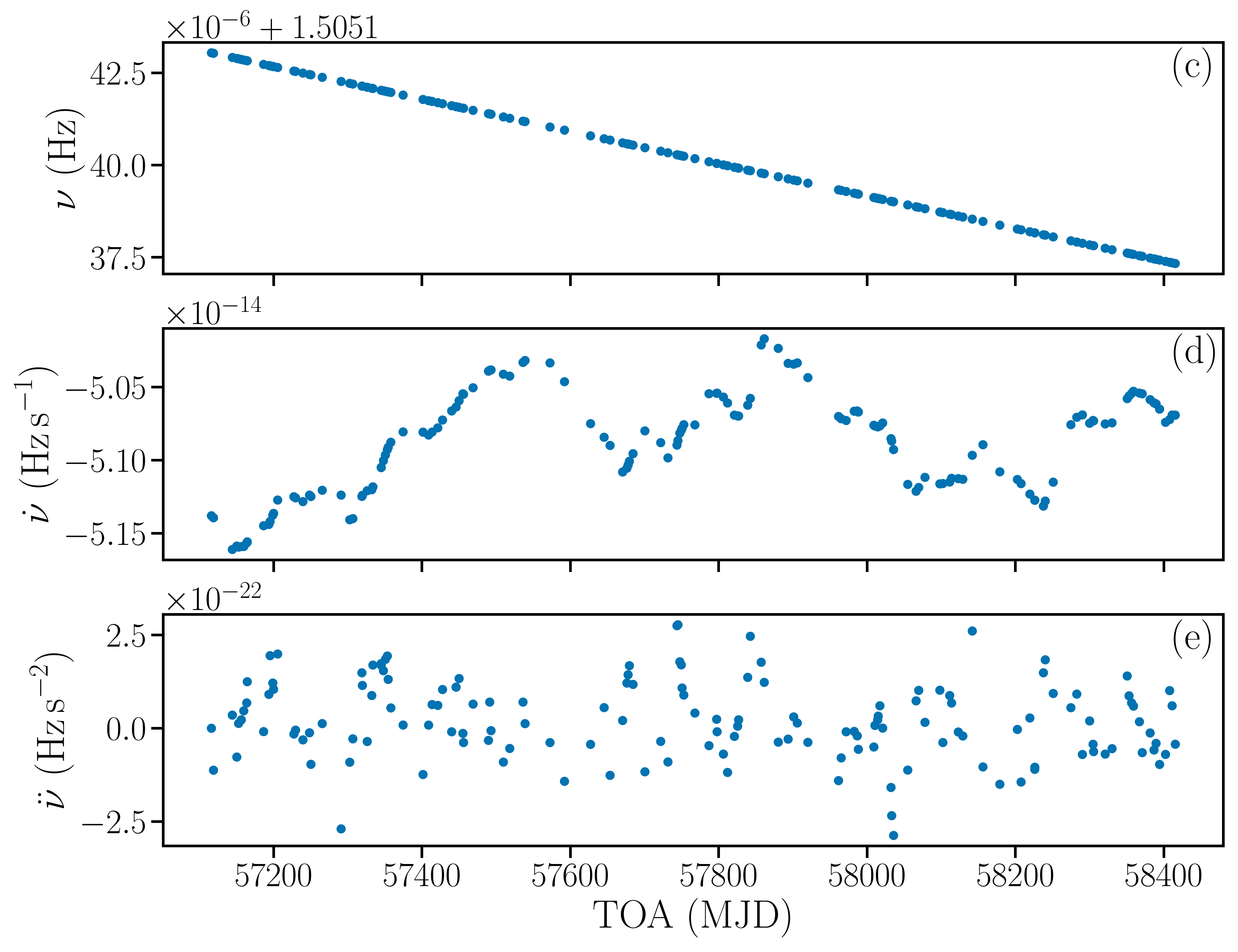}
 \caption{Actual and synthetic rotational evolution of the representative object PSR J0942$-$5552. (a) Actual phase residuals (units: s) versus observing epoch (units: MJD) taken from~\protect\cite{LowerBailes2020}. (b) Synthetic phase residuals generated by solving (\ref{Eq_SecII:model})--(\ref{Eq:diff_braking_law}) with the injected parameters in Table~\ref{Table_subsecII:example_injected_values}. The actual and synthetic phase residuals resemble each other visually. (c) Spin frequency $\nu(t)$ (units: Hz) versus observing epoch. (d) $\dot{\nu}(t)$ (units: ${\rm Hz\,s}^{-1}$) versus observing epoch. (e) $\ddot{\nu}(t)$ (units: ${\rm Hz\,s}^{-2}$) versus observing epoch. Note the powers of 10 defining the scales at the top left of panels (c), (d), and (e). Panels (c), (d), and (e) are obtained by solving Equations (\ref{Eq_SecII:model})--(\ref{Eq:diff_braking_law}) with the values listed in Table~\ref{Table_subsecII:example_injected_values}. We set the uncertainties for $\nu, \dot{\nu}, \ddot{\nu}$ following the procedure detailed in Section~\ref{subsecII:Bayesian_param_est_tnest}. The fractional fluctuations in panels (c), (d), and (e) are of order $\sim 10^{-8}$, $\sim 10^{-3}$, and $\sim 1$ respectively, in line with the actual data from PSR J0942$-$5552.}
\label{fig_subsecII:example_f2_walk}
\end{figure}

To guide the reader in reproducing the results in this paper, we start with a short worked example. We create a single random realization of synthetic data by solving (\ref{Eq_SecII:model})--(\ref{Eq:diff_braking_law}) for the injected parameters in Table~\ref{Table_subsecII:example_injected_values}, which emulate the arbitrary but representative object PSR J0942$-$5552 \citep{LowerBailes2020}. The object is chosen because its measured braking index satisfies $n = \nu(t_0) \ddot{\nu}(t_0) / \dot{\nu}(t_0)^2 = 4591.4$, i.e. its observed rotational evolution is dominated by some process other than secular electromagnetic braking with $n=n_{\rm pl}\approx3$. The timing noise amplitude $\sigma_{\ddot{\nu}} = 2.5\times10^{-50}~{\rm Hz}^{2}{\rm s}^{-5}$ is set in order to make the synthetic phase residuals resemble qualitatively the real phase residuals. In this example, the Brownian model (\ref{Eq_SecII:model})--(\ref{Eq_subsecII:memory_less}) driving the spin fluctuations differs from the red noise model (\ref{Eq:Temponest_TimingNoise}) assumed in the timing analysis.

 In Figure~\ref{fig_subsecII:example_f2_walk}, we compare visually the synthetic rotational evolution with the actual evolution observed by~\cite{LowerBailes2020} in PSR J0942$-$5552. The aim of the comparison is to check that solving (\ref{Eq_SecII:model})--(\ref{Eq_subsecII:memory_less}) produces phase residuals and associated rotational dynamics which are realistic astrophysically, in the sense that they resemble the actual observed data from a representative pulsar. The top two panels, viz. Figures~\ref{fig_subsecII:example_f2_walk}(a) and~\ref{fig_subsecII:example_f2_walk}(b), present the actual and synthetic phase residuals respectively. It is clear visually that the fluctuations in the two plots have similar amplitudes and wandering time-scales. The lower three panels, viz. Figures ~\ref{fig_subsecII:example_f2_walk}(c)--\ref{fig_subsecII:example_f2_walk}(e), display the synthetic evolution of $\nu(t)$, $\dot{\nu}(t)$, and $\ddot{\nu}(t)$ respectively as functions of observing epoch $t$ (in units of MJD). The results are in accord qualitatively with the output of~\tempoDOS, i.e. $\nu(t)$ is dominated by approximately linear spin down [with $T_{\rm obs} \ll \nu/(2\dot{\nu})$], $\dot{\nu}$ is approximately constant with fluctuations of fractional amplitude $\sim 10^{-3}$, and $\ddot{\nu}$ fluctuates appreciably about its secular electromagnetic value $\langle \ddot{\nu} \rangle = 5.262\times10^{-27} {\rm Hz\,s}^{-2}$ (corresponding to $n_{\rm pl}=3$) with fluctuations of fractional amplitude $\sim 1$. These properties are characteristic qualitatively of many of the pulsars timed by~\cite{LowerBailes2020}. Note that the small fluctuations in $\nu(t)$ and $\dot{\nu}(t)$ are inherited from $\ddot{\nu}(t)$ by integrating the Wiener increments $d{\bf B}(t)$ (with $\sigma_{\ddot{\nu}}^{2} \neq 0$) in (\ref{Eq_SecII:model}), as discussed in Section~\ref{subsecII:Noisy_ts_gen}, and the larger fluctuations in $\ddot{\nu}(t)$ overwhelm the secular $n_{\rm pl}=3$ trend in Figure~\ref{fig_subsecII:example_f2_walk}(e).

We feed the synthetic TOAs in Figure~\ref{fig_subsecII:example_f2_walk}(b) into \tempoDOS~to estimate $
\theta$ and $\Delta \theta$ in preparation for selecting \temponest~priors according to the recipe in Section~\ref{subsecII:Bayesian_param_est_tnest}. \tempoDOS~demands an initial guess for the ephemeris, which we take to be ${\bf X}(t_{0})$ from Table~\ref{Table_subsecII:example_injected_values}. For the synthetic data in Figure~\ref{fig_subsecII:example_f2_walk}(b), \tempoDOS~returns ${\rm RA}=2.54051~{\rm rad}, {\rm DEC}=-0.97532~{\rm rad}, \nu=1.5051~{\rm Hz}, \dot{\nu}=-5.0586\times10^{-14}~{\rm Hz\,s}^{-1}$ and $\ddot{\nu}=-2.38\times10^{-24}~{\rm Hz\,s}^{-2}$, with uncertainties $\Delta{\rm RA}=5.846\times10^{-6}~{\rm rad},\Delta{\rm DEC}=3.346\times10^{-6}~{\rm rad}, \Delta\nu=4.7704\times10^{-10}~{\rm Hz}, \Delta \dot{\nu}=1.9309\times10^{-17}~{\rm Hz\,s}^{-1}$ and $\Delta \ddot{\nu}=-3.2965\times10^{-25}~{\rm Hz\,s}^{-2}$, implying the priors summarized in Table~\ref{Table:tempo2_params_table}. Except for $\ddot{\nu}$, which is overestimated by $\sim10^{3}$, \tempoDOS's estimates for ${\rm RA},{\rm DEC},\nu$, and $\dot{\nu}$ are close to the injected values, with fractional differences ranging from $10^{-8}$ to $10^{-2}$.

\begin{table}
\caption{Prior ranges set from \tempoDOS~estimates of $\theta$ and $\Delta \theta$ for the worked example in Section~\ref{subsecII:gen_data_get_n_example}.}
\flushleft
\begin{tabular}{p{2cm}p{2.5cm}p{2.5cm}}
\hline
Parameter & Units & Prior range \\
\hline
RA & ${\rm rad}$ & $(2.54050,2.54052)$ \\
DEC & ${\rm rad}$ & $(-0.9753,-0.9751)$ \\
 $\nu$ & ${\rm Hz}$ &  $(1.5050,1.5052)$ \\
$\dot{\nu}$ & ${\rm Hz\,s}^{-1}$ & $(-10^{-12},0)$\\
$\ddot{\nu}$ & $10^{-22}~{\rm Hz\,s}^{-2}$ & $(-3.3,3.3)$\\
\hline
\end{tabular}
\label{Table:tempo2_params_table}
\end{table}

 We feed into \temponest~the synthetic TOAs from Figure~\ref{fig_subsecII:example_f2_walk}(b), as well as the \tempoDOS~estimates of $\theta$ and the priors in Table~\ref{Table:tempo2_params_table}, to obtain the final estimates for $\theta$, $\Delta \theta, A_{\rm red}$, and $\beta$. For this trial the inferred noise parameters are $\log A_{\rm red}=-9.05\pm0.1$, and $\beta=6.22\pm0.4$. These are consistent with the reported values for PSR J0942$-$5552 in \cite{LowerBailes2020}, namely $\log A_{\rm red}=-9.03\pm0.2$, and $\beta=5.88^{+1.6}_{-1.1}$. The values recovered by \temponest~for $\nu, \dot \nu,~{\rm and}~\ddot \nu$ are recorded in the right-hand column of Table~\ref{Table_subsecII:example_injected_values}. The fractional differences between the recovered and injected values are $\sim 10^{-9},\sim 10^{-4}$, and $~10^{2}$ for $\nu,\dot{\nu}$, and $\ddot{\nu}$ respectively. The recovered parameters in the right column of Table~\ref{Table_subsecII:example_injected_values}, corresponding to the peak of the posterior, are combined through (\ref{Eq:Intro_n}) to yield a synthetic measurement of the braking index, viz. $ n \pm \Delta n = 5248 \pm 1118$. Here $\Delta n$ represents the formal uncertainty returned by \temponest~when estimating $\Delta \ddot{\nu}=1.961\times10^{-24}~{\rm Hz\,s}^{-2}$. The measurement $n\pm \Delta n$ is consistent with the actual astrophysical value $n=4591^{+3.1}_{-3.5}$ for PSR J0942$-$5552 reported by~\cite{LowerBailes2020}. This is encouraging, when one recalls that the synthetic data are constructed to emulate PSR J0942$-$5552 and its timing noise parameters. 

In the above example, synthetic data are created following the braking law (\ref{Eq:diff_braking_law}) with $n_{\rm pl}=3$. Yet timing noise with amplitude $\sigma_{\ddot{\nu}}^{2} = 2.5\times10^{-50} {\rm Hz}^{2}\,{\rm s}^{-5}$, which matches the observed phase residuals qualitatively and visually [see Figures~\ref{fig_subsecII:example_f2_walk}(a) and \ref{fig_subsecII:example_f2_walk}(b)], masks the secular evolution of $\ddot{\nu}$. The synthetic measurement satisfies $n \gg n_{\rm pl}$. The measured $\Delta\ddot{\nu}$ value is such that the range $n\pm \Delta n$ excludes $n_{\rm pl}$ artificially. In other words, the representative example emulating PSR J0942$-$5552 demonstrates one instance of the central point of the paper, namely that timing noise can corrupt the inference of $n$ in some pulsars, when the timing noise amplitude takes typical astrophysical values. Having demonstrated the point in principle in one specific example, the logical next step is to test other pulsars with $n \gg n_{\rm pl}$, e.g.\ some of the objects studied by~\cite{LowerBailes2020}. As there is no way at present to measure $n_{\rm pl}$ independently in any real pulsar, we perform Monte Carlo numerical experiments with synthetic data, following the recipe in Section~\ref{subsecII:gen_data_get_n_example}, to characterize the statistical properties of the recovered braking indices 
 as a function of the injected noise strength $\sigma_{\ddot \nu}^{2}~$. These tests and their results form the basis of Sections~\ref{sec:Recovered_n_stats} and~\ref{sec:Disp_n_vs_noise}.

\section{Statistics of the measured braking index}
\label{sec:Recovered_n_stats}

\begin{figure}
\flushleft
 \includegraphics[width=\columnwidth]{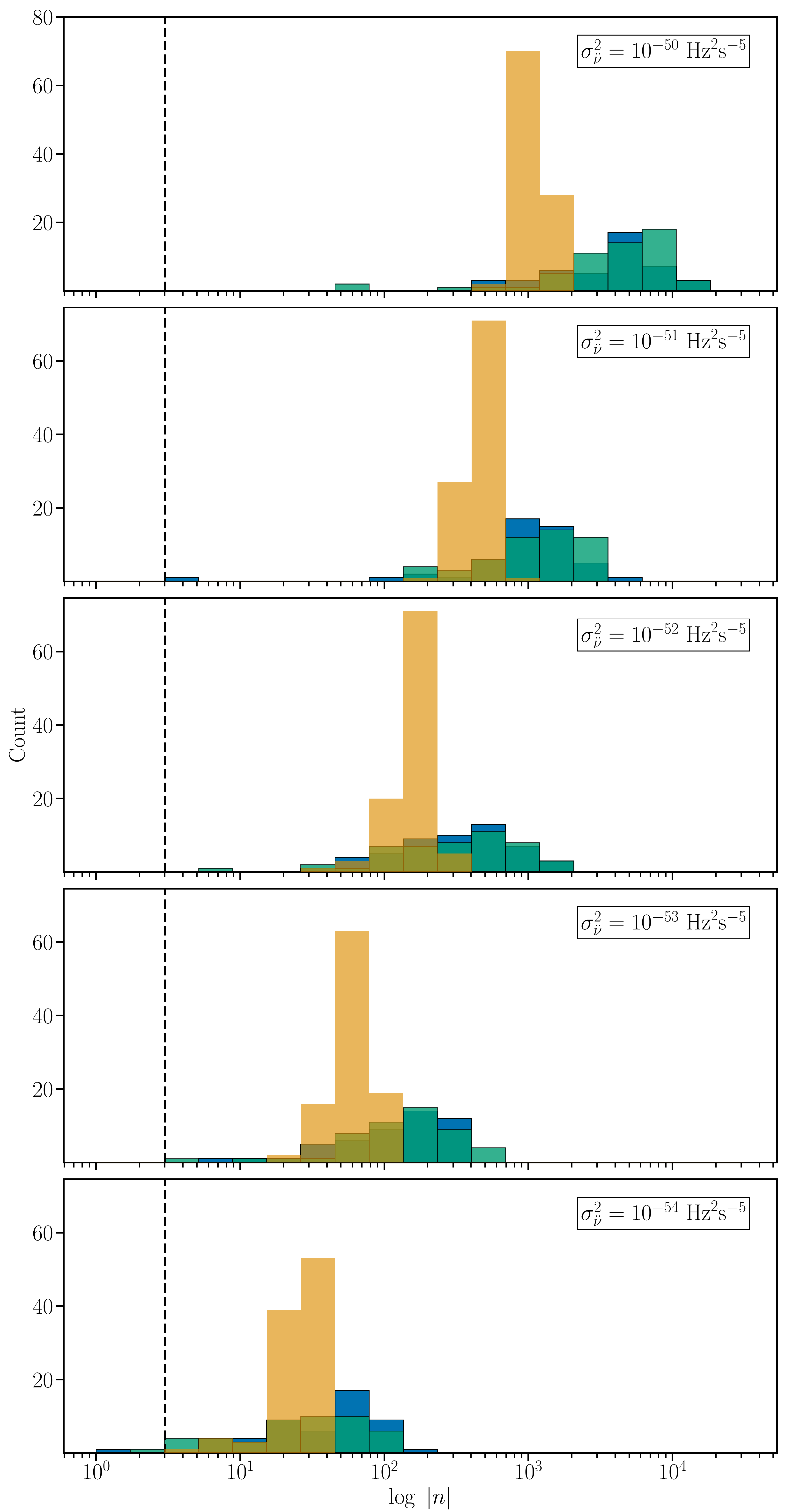}
 \caption{Distribution of $n$ measurements (blue histograms for $n>0$, cyan histograms for $n<0$) and their formal uncertainties $\Delta n$ (orange histograms) reported by~\temponest~through (\ref{Eq:Intro_n}) and (\ref{Eq:Uncertainty_n}) for astrophysically plausible timing noise amplitudes running from $\sigma_{\ddot{\nu}}^2 = 10^{-54}~{\rm Hz}^{2}{\rm s}^{5}$ in the bottom panel to $\sigma_{\ddot{\nu}}^2 = 10^{-50}~{\rm Hz}^{2}{\rm s}^{5}$ in the top panel (see legend). All panels are constructed from $10^2$ random realizations of synthetic data at the given $\sigma^{2}_{\ddot{\nu}}$ and spin parameters characteristic of PSR J0942$-$5552 from Table~\ref{Table_subsecII:example_injected_values}, noting that the results are approximately independent of $\nu(t_{0})$ and $\dot{\nu}(t_{0})$ as discussed in Section~\ref{subsec_IV:DISP_n_vs_f_fdot}. The black dotted line represents the injected value $n_{\rm pl}=3$. In the bottom panel  the averages of $n$ and $\Delta n$ are $8$ and $26.63$ respectively, with the measurements satisfying $n-\Delta n \leq n_{\rm pl} \leq n+\Delta n$ in 41 out of 100 trials. In the top panel, the averages of $n$ and $\Delta n$ are $-798.73$ and $1112.58$ respectively, with the measurements satisfying $n-\Delta n \leq n_{\rm pl} \leq n+\Delta n$ in 8 out of 100 trials. The increase in ${\rm DISP}(n)$ from the bottom to the top panel is apparent from the rightward shift of the histograms along the logarithmic horizontal axis.}
\label{fig_subsecIII:hists_vs_sigmas}
\end{figure}

In this section we present the distribution of measured $n$ values returned by~\temponest~for a range of timing noise amplitudes $\sigma_{\ddot{\nu}}^2$ ranging from relatively low to relatively high, as specified below. The goal is to determine under what conditions $n_{\rm pl}$ is measured reliably via $n$, i.e.\ for what values of $\sigma_{\ddot{\nu}}^2$ one obtains $n\approx n_{\rm pl}$ (low noise) as opposed to $ |n | \gg n_{\rm pl}$ (high noise).

To meet the above goal, we repeat the procedure presented in Section~\ref{subsecII:gen_data_get_n_example} using $100$ random realizations of synthetic data per $\sigma^{2}_{\ddot{\nu}}$, as opposed to a single random realization. All the realizations use the same parameters as in Table~\ref{Table_subsecII:example_injected_values} except for the injected noise strength, which covers the range $10^{-54} \leq \sigma^{2}_{\ddot{\nu}}/(1 \, {\rm Hz}^{2}{\rm s}^{-5} ) \leq 10^{-50}$. We select this range for $\sigma^{2}_{\ddot{\nu}}$, because it produces phase residuals consistent with those presented in \cite{LowerBailes2020} and \cite{ParthasarathyJohnston2020}. Specifically, $\sigma^{2}_{\ddot{\nu}}=10^{-54}~{\rm Hz}^{2}{\rm s}^{-5}$  and $\sigma^{2}_{\ddot{\nu}}=10^{-50}~{\rm Hz}^{2}{\rm s}^{-5}$ produce phase residuals of order $\sim 10^{-1}~{\rm ms}$ and $\sim10^{2}~{\rm ms}$, respectively. We set \temponest~priors using the same values as in Tables~\ref{Table:stochastic_params_table} and \ref{Table:tempo2_params_table}, except for $\ddot{\nu}$ whose priors are set following the recipe detailed in Section~\ref{subsecII:Bayesian_param_est_tnest}. We confirm below (see Section~\ref{subsec_IV:DISP_n_vs_f_fdot}) that the results for ${\rm DISP}(n)$ do not depend on the specific values of $\nu$ and $\dot{\nu}$. That is, the results for ${\rm DISP}(n)$ obtained for the injection parameters in Table~\ref{Table_subsecII:example_injected_values} (with only $\ddot{\nu}$ and $\sigma_{\ddot{\nu}}^{2}$ varying) transfer approximately unchanged to pulsars with arbitrary $\nu$ and $\dot{\nu}$, e.g.\ in the ATNF database.        

Figure~\ref{fig_subsecIII:hists_vs_sigmas} demonstrates how the $n$ value measured from (\ref{Eq:Intro_n}) becomes less reliable as an estimate of $n_{\rm pl}$, as the timing noise amplitude $\sigma_{\ddot{\nu}}^2$ increases from the bottom panel ($\sigma^{2}_{\ddot{\nu}}=10^{-54}~{\rm Hz}^{2}{\rm s}^{-5}$) to the top panel ($\sigma^{2}_{\ddot{\nu}}=10^{-50}~{\rm Hz}^{2}{\rm s}^{-5}$). Each panel displays three histograms. The blue and cyan histograms bin and count the measured $n>0$ and $n<0$ values, respectively, as computed from the~\temponest~output substituted into (\ref{Eq:Intro_n}). The orange histogram bins and counts the formal uncertainty $\Delta n$ reported automatically by~\temponest~via (\ref{Eq:Uncertainty_n}). The histograms are plotted on a logarithmic scale, i.e. $\log_{10} \vert n \vert$, for clarity. We observe two things. First, the orange histograms are narrower than the blue and cyan histograms for $\sigma_{\ddot{\nu}}^2 \gtrsim 10^{-54}~{\rm Hz}^{2}{\rm s}^{-5}$. For example, the full widths half maximum (FWHMs) of the orange and the summed blue and cyan histograms are $16.11$ and $89.36$ respectively at $\sigma_{\ddot{\nu}}^2 = 10^{-54}~{\rm Hz}^{2}{\rm s}^{-5}$ and rise to $235.38$ and $14932$ respectively at $\sigma_{\ddot{\nu}}^2 = 10^{-50}~{\rm Hz}^{2}{\rm s}^{-5}$. This means that the formal uncertainty on $n$, as calculated by (\ref{Eq:Uncertainty_n}), underestimates the dispersion in $n$ values associated with different random realizations of the timing noise, bearing in mind that timing observations of a real pulsar sample one out of the ensemble of possible realizations, and there is no way to know or predict which one. Second, the FWHMs of the summed blue and cyan histograms are wider than $n_{\rm pl}$ under a range of conditions. For example, the FWHM rises from $29.78\,n_{\rm pl}$ at $\sigma_{\ddot{\nu}}^2 = 10^{-54}~{\rm Hz}^{2}{\rm s}^{-5}$ to $4977\,n_{\rm pl}$ at $\sigma_{\ddot{\nu}}^2 = 10^{-50}~{\rm Hz}^{2}{\rm s}^{-5}$. Similarly, ${\rm ERR}(n)$ rises from ${\rm ERR}(n)=-17.61$ at $\sigma_{\ddot{\nu}}^2 = 10^{-54}~{\rm Hz}^{2}{\rm s}^{-5}$ to ${\rm ERR}(n)=-801.73$ at $\sigma_{\ddot{\nu}}^2 = 10^{-50}~{\rm Hz}^{2}{\rm s}^{-5}$. Indeed, \temponest~routinely returns $n$ values of either sign for $\sigma_{\ddot{\nu}}^2 \gtrsim 10^{-54}~{\rm Hz}^{2}{\rm s}^{-5}$, in line with many observational studies \citep{JohnstonGalloway1999,ChukwudeChidiOdo2016,ParthasarathyJohnston2020,LowerBailes2020}. This means that the $n$ measurements generated by combining the~\temponest~output with (\ref{Eq:Intro_n}) are unreliable measurements of $n_{\rm pl}$ for $\sigma_{\ddot{\nu}}^2 \gtrsim 10^{-54}~{\rm Hz}^{2}{\rm s}^{-5}$, noting again that it is impossible to know or predict which random noise realization out of the ensemble of possibilities is sampled by timing observations of any particular real pulsar. 

Let us look at the results in Figure~\ref{fig_subsecIII:hists_vs_sigmas} in another, equivalent way. As $\sigma_{\ddot{\nu}}^2$ increases, fewer of the $10^2$ individual trial measurements in the ensemble agree with $n_{\rm pl}$ within the formal uncertainty reported by~\temponest. In the bottom panel, for example, only 41 out of 100 trials satisfy $n- \Delta n \leq n_{\rm pl} \leq n + \Delta n$. For the rest of the panels, in ascending order, only 18, 27, 17, and 8 of the 100 trials satisfy $n- \Delta n \leq n_{\rm pl} \leq n + \Delta n$. \footnote{The non-monotonic order for $\sigma = 10^{-54}~{\rm Hz}^{2}{\rm s}^{-5}$ and $10^{-53}~{\rm Hz}^{2}{\rm s}^{-5}$ seems to be a statistical fluctuation. It may be checked using an ensemble containing more than 100 trials, when sufficient computational resources become available.} In other words, as the timing noise amplitude increases, the confidence interval $n \pm \Delta n$ reported by~\temponest~excludes $n_{\rm pl}$ in more realizations. This result is not surprising. A random noise process corrupts the measurement of an underlying secular trend if it is strong enough; the question is how strong, and the answer here is $\sigma_{\ddot{\nu}}^2 \gtrsim 10^{-54}~{\rm Hz}^{2}{\rm s}^{-5}$ for the parameters in Table~\ref{Table_subsecII:example_injected_values}, which are broadly characteristic of known pulsars, e.g.\ objects in the ATNF database \citep{LowerBailes2020,ParthasarathyJohnston2020}. For example, the top panel in Figure~\ref{fig_subsecIII:hists_vs_sigmas} is not extreme. It represents an astrophysically motivated timing noise amplitude, viz. $\sigma^{2}_{\ddot{\nu}}=10^{-50}~{\rm Hz}^{2}{\rm s}^{-5}$, which produces phase residuals whose variations are comparable to those presented in Figure~2 of \cite{LowerBailes2020}, Figure~4 of \cite{ParthasarathyJohnston2020} or Figures~\ref{fig_subsecII:example_f2_walk}(a) and~\ref{fig_subsecII:example_f2_walk}(b) in this paper. The histogram in this panel yields $\langle n \rangle=-798.73$, ${\rm max}(n)=17163$, ${\rm ERR}(n)=-801.73$, and ${\rm max}(\Delta n)=1655.8$. 

 It may be argued that individual trial measurements of $n$ in the histograms in Figure~\ref{fig_subsecIII:hists_vs_sigmas} are scattered widely relative to $n_{\rm pl}$, but the peak of the histogram matches $n_{\rm pl}$ accurately. There are two responses to this argument. First, one never has the opportunity to construct the blue and cyan histograms in Figure~\ref{fig_subsecIII:hists_vs_sigmas} when observing a real pulsar. One observes a single random realization of the timing noise -- the actual realization -- without any way to predict where it lies within the histogram (and therefore where it lies relative to $n_{\rm pl}$). If there are physical reasons to think that $n_{\rm pl}$ is the same in every pulsar, one could construct the blue and cyan histograms in principle by measuring $n$ in different pulsars. However, existing measurements of nonanomalous braking indices in low-noise pulsars, e.g.\ the Crab pulsar or PSR J1640$-$4631, suggest that $n_{\rm pl}$ satisfies $1\leq n_{\rm pl} \leq 3$ and may conceivably span an even wider range~\citep{LyneAG1993,ArchibaldGotthelf2016}.  Another possible strategy is to construct the two-dimensional histogram in the $n$--$\sigma_{\ddot{\nu}}^{2}$ plane for multiple observed pulsars. This strategy might shed some light on the distribution in a population sense. We expand on this point in Sections~\ref{subsec_IV:DISP_n_vs_f_fdot} and~\ref{sec:Conclusions}. The second response is that even the peak of the summed blue and cyan histograms does not match $n_{\rm pl}$ accurately. For example, in the top panel of Figure~\ref{fig_subsecIII:hists_vs_sigmas}, with $\sigma_{\ddot{\nu}}^2 =10^{-50}~{\rm Hz}^{2}{\rm s}^{-5}$, the mean and median of the $10^2$ trials are $-798.73$ and $-1561.9$ respectively, compared to $n_{\rm pl}=3$. Generally speaking,~\temponest~overestimates $\vert \ddot{\nu} \vert$ on average over the ensemble of trials and underestimates $\Delta\ddot{\nu}$.

\section{Dispersion of measured braking indices versus timing noise properties}
\label{sec:Disp_n_vs_noise}

\begin{figure}
\flushleft
 \includegraphics[width=\columnwidth]{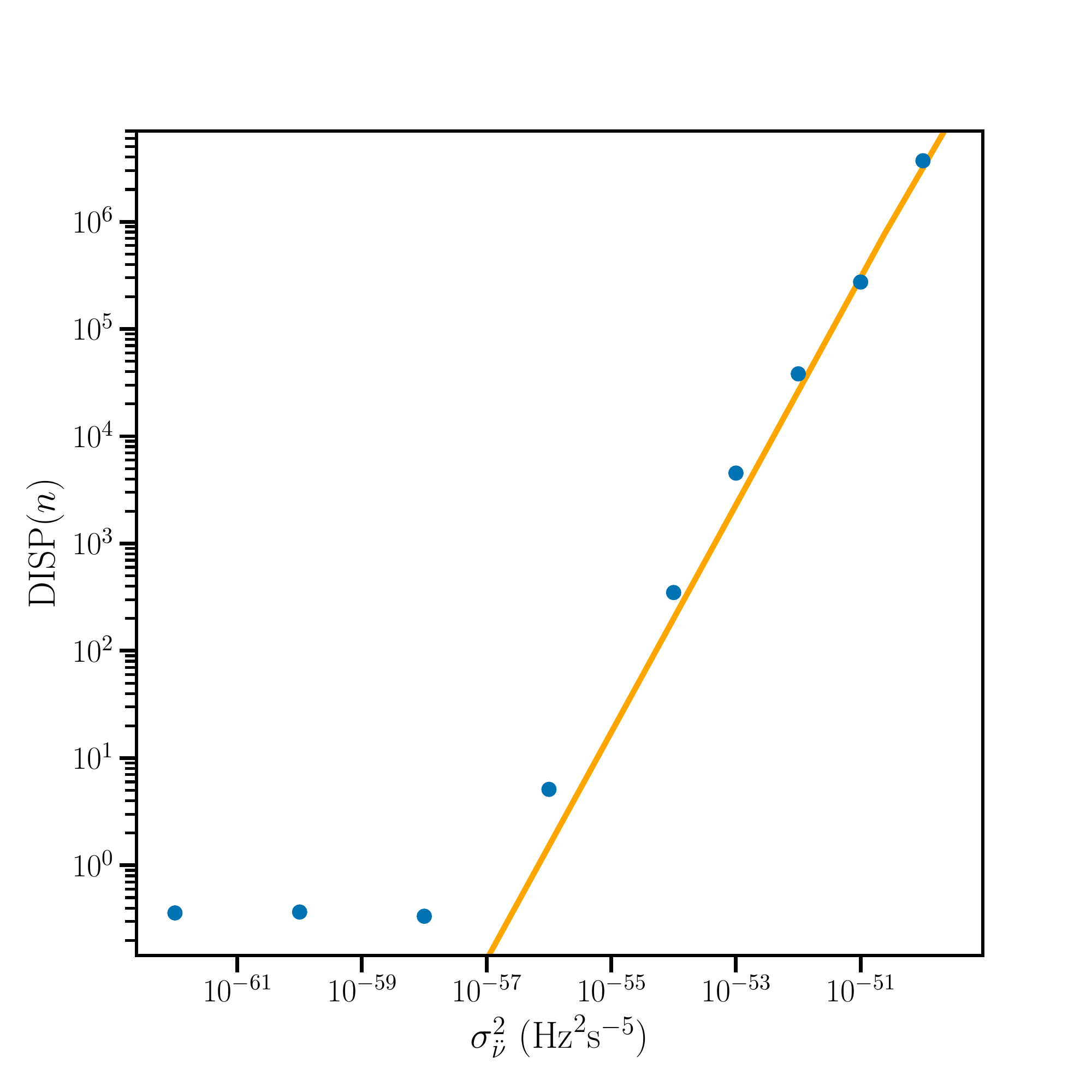}
 \caption{Dispersion ${\rm DISP}(n)$ (vertical axis) versus $\sigma^{2}_{\ddot{\nu}}$ (horizontal axis). Each point represent ${\rm DISP}(n)$ for $100$ synthetic data realizations. We repeat the procedure detailed in Section~\ref{subsecII:gen_data_get_n_example} to obtain synthetic $n$ measurements to calculate ${\rm DISP}(n)$. The thick solid line represents ${\rm DISP}(n)$ as predicted by the analytic theory in Appendix~\ref{Appendix:Theory_anom_n}, when $\ddot{\nu}(t)$ is measured nonlocally by finite differencing the time series $\dot{\nu}(t)$. The orange line is a falsifiable prediction from first principles; it is not a fit. For low timing noise strengths (the leftmost three points) we find ${\rm DISP}(n)\approx 0.4$ and more than $90\%$ of the recovered $n\pm\Delta n$ intervals include $n_{\rm pl}$, where $\Delta n$ is the formal uncertainty returned by~\temponest~via (\ref{Eq:Uncertainty_n}). ${\rm DISP}(n)$ grows approximately quadratically with $\sigma_{\ddot{\nu}}$ for $\sigma^{2}_{\ddot{\nu}}\gtrsim 10^{-57}~{\rm Hz}^{2}{\rm s}^{-5}$. For $\sigma^{2}_{\ddot{\nu}}=10^{-50}~{\rm Hz}^{2}{\rm s}^{-5}$ only $8\%$ of the trials yield $n\pm\Delta n$ intervals that include $n_{\rm pl}$. The formal uncertainty interval $n\pm\Delta n$ excludes $n_{\rm pl}$ more often, as $\sigma^{2}_{\ddot{\nu}}$ increases.}
\label{fig_subsecIV:DISP_vs_sigmas}
\end{figure}

In this section we calculate the spread of $n$ measurements as a function of $\sigma^{2}_{\ddot{\nu}}$. The aim is to quantify the condition on the noise amplitude that produces $\vert n \vert \gg n_{\rm pl}$ in a probabilistic sense, when one noise realization out of a random ensemble is measured, as in real pulsar timing experiments. To this end, Section~\ref{subsec_IV:DISP_n_vs_sigma2} calculates the dispersion ${\rm DISP}(n)$ versus $\sigma^{2}_{\ddot{\nu}}$, using synthetic data with PSR J0942$-$5552-like values, to find a condition on $\sigma^{2}_{\ddot{\nu}}$ that yields ${\rm DISP}(n) \gg 1$. Section~\ref{subsec_IV:DISP_n_vs_TN_PSD} calculates ${\rm DISP}(n)$ as a function of the amplitude and shape of the PSD phase residuals, i.e.\ the parameters $A_{\rm red}$ and $\beta$ respectively in equation~(\ref{Eq:Temponest_TimingNoise}). For the sake of completeness, Section~\ref{subsec_IV:ERRn_vs_sigma2} calculates the bias ${\rm ERR}(n)$ versus $\sigma^{2}_{\ddot{\nu}}$, even though ${\rm ERR}(n)$ is not relevant directly to astronomical measurements for the reasons given following equation (\ref{Eq:Uncertainty_n}). Section~\ref{subsec_IV:DISP_n_vs_f_fdot} calculates ${\rm DISP}(n)$ for various values of $\nu(t_{0})$ and $\dot{\nu}(t_{0})$, holding $\ddot{\nu}(t_{0})$ and $n_{\rm pl}$ fixed, to verify that ${\rm DISP}(n)$ does not depend strongly on $\nu(t_{0})$ and $\dot{\nu}(t_{0})$. Hence the results in this paper can be applied to any object in the ATNF pulsar database~\cite{ManchesterHobbs2005} without adjusting significantly for $\nu(t_{0})$ and $\dot{\nu}(t_{0})$.    

\subsection{${\rm DISP}(n)$ versus $\sigma^{2}_{\ddot{\nu}}$}
\label{subsec_IV:DISP_n_vs_sigma2}

To explore ${\rm DISP}(n)$ versus $\sigma^{2}_{\ddot{\nu}}$, we apply the same numerical recipe outlined in Section~\ref{sec:Recovered_n_stats} to the extended range $10^{-62} \leq \sigma^{2}_{\ddot{\nu}} / (1\,{\rm Hz}^{2}{\rm s}^{-5}) \leq 10^{-50}$. Figure~\ref{fig_subsecIV:DISP_vs_sigmas} displays ${\rm DISP}(n)$ (vertical axis), calculated using (\ref{Eq:Dispersion_n}) and $n$ measured from $100$ synthetic data realizations, versus the timing noise amplitude, $\sigma^{2}_{\ddot{\nu}}$. The blue dots are obtained from numerical simulations; the orange line is a theoretical prediction discussed further below in this section and Appendix~\ref{Appendix:Theory_anom_n}. Figure~\ref{fig_subsecIV:DISP_vs_sigmas} reveals two distinct regimes: (i) a flat region with ${\rm DISP}(n) \approx 0.4$, spanning $10^{-62} \leq \sigma^{2}_{\ddot{\nu}} / (1\,{\rm Hz}^{2}{\rm s}^{-5}) \leq 10^{-58}$, and (ii) a linear region with $0.4 \leq {\rm DISP}(n) \leq 4\times10^{6}$, spanning $10^{-58} \leq \sigma^{2}_{\ddot{\nu}} / (1\,{\rm Hz}^{2}{\rm s}^{-5}) \leq 10^{-50}$. 

In region (i) in Figure~\ref{fig_subsecIV:DISP_vs_sigmas}, \temponest~recovers accurately the secular value  of $\ddot{\nu}(t_{0})$, as well as $\nu(t_{0})$ and $\dot{\nu}(t_{0})$ of course, and therefore returns a reliable estimate of the secular braking index $n_{\rm pl}$ via (\ref{Eq:Intro_n}). The number of measured $n\pm\Delta n$ uncertainty intervals that include $n_{\rm pl}$ are $95,97$, and $93$ out of $100$, for $\sigma^{2}_{\ddot{\nu}}$ values of $10^{-62}~{\rm Hz}^{2}{\rm s}^{-5}, 10^{-60}~{\rm Hz}^{2}{\rm s}^{-5}$, and $10^{-58}~{\rm Hz}^{2}{\rm s}^{-5}$ respectively. In other words, the evolution of $\dot{\nu}$ is predominantly secular $[\dot{\nu}(t)\approx\dot{\nu}_{\rm em}(t)]$ with negligible contributions from the noise in $\ddot{\nu}$. For comparison, in region (i), $\dot{\nu}$ has fluctuations of fractional amplitude $\sim 10^{-6}$, while in Section~\ref{subsecII:gen_data_get_n_example} these fluctuations are $10^{3}$ times bigger.

In region (ii) in Figure~\ref{fig_subsecIV:DISP_vs_sigmas}, where we find ${\rm DISP}(n)\propto \sigma^{2}_{\ddot{\nu}}$, the spread (standard deviation) of measured $n$ values spans $ 7.5\times10^{-1} \leq  ( \langle n^2 \rangle - n_{\rm pl}^2)^{1/2} \leq 6.4\times10^{2}$ for the blue dots. As expected, higher timing noise creates a greater spread in $n$, with $\vert n \vert \geq n_{\rm pl}$ for $\sigma^{2}_{\ddot{\nu}} \geq 10^{-54}~{\rm Hz}^{2}{\rm s}^{-5}$. In region (ii), the number of measured $n\pm\Delta n$ uncertainty intervals that include $n_{\rm pl}$ are $50,41,18,27,17$, and $8$ out of $100$, for $\sigma^{2}_{\ddot{\nu}} / (1\,{\rm Hz}^{2}{\rm s}^{-5})$ equaling $10^{-56}, 10^{-54},10^{-53},10^{-52},10^{-51}$, and $10^{-50}$ respectively. In region (ii), $\ddot{\nu}$ exhibits fluctuations of fractional amplitude $\sim 1$, and the evolution of $\dot{\nu}(t)$ is mainly driven by the noise term, viz. $\dot{\nu} \approx \int dt'\,\ddot{\nu}(t')$, instead of $\ddot{\nu}_{\rm em}(t)$. Note that the demarcation point between regions (i) and (ii) depends on exactly how the analyst wishes to define a reliable measurement of $n_{\rm pl}$. Is it that ${\rm DISP}(n)$ falls below some analyst-selected threshold, or is it that the probability of $n\pm\Delta n$ including $n_{\rm pl}$ exceeds some threshold? Both criteria are related and similar, but they are not exactly the same. For instance, $\sigma^{2}_{\ddot{\nu}}=10^{-56}~{\rm Hz}^{2}{\rm s}^{-5}$ yields $[\ddot{\nu}(t)-\ddot{\nu}_{\rm em}(t)]/\ddot{\nu}=0.77, {\rm DISP}(n)=5.11$, and ${\rm max}(n)=19.15$, but only half of the measured $n\pm\Delta n$ intervals contain $n_{\rm pl}$. By contrast, $\sigma^{2}_{\ddot{\nu}}=10^{-58}~{\rm Hz}^{2}{\rm s}^{-5}$ yields $[\ddot{\nu}(t)-\ddot{\nu}_{\rm em}(t)]/\ddot{\nu}=0.58, {\rm DISP}(n)=0.34$, and ${\rm max}(n)=7.75$, but $93$ out of the $100$ measured $n\pm\Delta n$ intervals contain $n_{\rm pl}$.

One might wonder whether it is possible to predict the scaling in Figure~\ref{fig_subsecIV:DISP_vs_sigmas} theoretically. The answer is yes. In Appendix~\ref{Appendix:Theory_anom_n} the Brownian model (\ref{Eq_SecII:model})--(\ref{Eq_subsecII:memory_less}) is solved analytically in the regime where one has $\nu(t) \approx \nu_{\rm em}(t)$ but $\ddot{\nu}(t)$ fluctuates appreciably about $\ddot{\nu}_{\rm em}(t)$, which is relevant to pulsar timing experiments. The solutions are applied to calculate ${\rm DISP}(n)$ via~(\ref{Eq:Dispersion_n}) and the result is

\begin{equation}
    {\rm DISP}(n) = \frac{\sigma^{2}_{\ddot{\nu}}}{\gamma_{\ddot{\nu}}^{2}\ddot{\nu}^{2}_{\rm em}(t_{0})T_{\rm obs}}.
    \label{Eq_SecIV:DISP_n_vs_sigma_gammaT}
\end{equation}

Equation (\ref{Eq_SecIV:DISP_n_vs_sigma_gammaT}) is a central result of the paper. It is drawn as the orange, diagonal line in Figure~\ref{fig_subsecIV:DISP_vs_sigmas}. We emphasize that it is not a best fit, nor is it a phenomenological formula. Rather, it is a falsifiable, analytic result derived from first principles by solving (\ref{Eq_SecII:model})--(\ref{Eq_subsecII:memory_less}) and calculating $\ddot{\nu}$ and hence $n$ by finite differencing the time series $\dot{\nu}(t)$. (Alternative measurement strategies are discussed in Appendix~\ref{Appendix:Theory_anom_n}.) The agreement between the blue dots and orange line in Figure~\ref{fig_subsecIV:DISP_vs_sigmas}, assuming the known, injected values of $\gamma_{\ddot{\nu}}$ and $\sigma_{\ddot{\nu}}$, is encouraging and confirms the validity of the synthetic measurement strategy involving~\temponest. Now that (\ref{Eq_SecIV:DISP_n_vs_sigma_gammaT}) is verified, it can be applied to real astronomical data, by substituting an estimate of $\sigma_{\ddot{\nu}}$ from~\temponest~measurements of $P_{\rm r}(f)$ (see Section~\ref{subsec_IV:DISP_n_vs_TN_PSD}), and assuming a fiducial value of $\gamma_{\ddot{\nu}} \sim 10^{-6} \, {\rm s^{-1}}$ motivated by pulsar glitch recovery time-scales or the autocorrelation time-scale of pulsar timing noise \citep{PriceLink2012,MeyersMelatos2021,MeyersO'Neill2021}, as well as $\ddot{\nu}_{\rm em}(t_0) \sim \dot{\nu}(t_0)^2/\nu(t_0)$. In other words, even without knowing $\gamma_{\ddot{\nu}}$ and $\sigma_{\ddot{\nu}}$ exactly a priori in a real pulsar, one can use (\ref{Eq_SecIV:DISP_n_vs_sigma_gammaT}) to predict approximately the degree to which timing noise masks $n_{\rm pl}$. 

Region (i) in Figure~\ref{fig_subsecIV:DISP_vs_sigmas} does not obey the trend in (\ref{Eq_SecIV:DISP_n_vs_sigma_gammaT}). The departure is expected. The theory in Appendix~\ref{Appendix:Theory_anom_n} does not describe the complicated instrumental uncertainties embedded in a~\temponest~measurement, e.g.\ quantified by $\Delta n$ and EFAC and EQUAD in Table~\ref{Table:stochastic_params_table}. In other words, in region (i), ${\rm DISP}(n)$ is dominated by instrumental uncertainties rather than the randomness of noise realizations. Fortunately, the breakdown of the theory in Appendix~\ref{Appendix:Theory_anom_n} does not matter in practical terms, because we find ${\rm DISP}(n) \lesssim 1$ and $n\approx n_{\rm pl}$ in region (i). That is, wherever instrumental uncertainties are the limiting factor,~\temponest~measurements of $n$ recover $n_{\rm pl} \approx n$ accurately anyway. 

\subsection{${\rm DISP}(n)$ versus~\temponest~phase residual spectrum}
\label{subsec_IV:DISP_n_vs_TN_PSD}

Equation (\ref{Eq_SecIV:DISP_n_vs_sigma_gammaT}) involves $\sigma_{\ddot{\nu}}^{2}$, whose exact value is unknown a priori in any real pulsar. However, it is possible to measure $\sigma_{\ddot{\nu}}^{2}$ approximately by relating it to the parameters $A_{\rm red}$ and $\beta$ in (\ref{Eq:Temponest_TimingNoise}) describing the amplitude and shape respectively of the phase residual PSD $P_{\rm r}(f)$ inferred by~\temponest~ \citep{LentatiAlexander2014,LentatiShannon2016,GoncharovReardon2021}.

\begin{figure}
\flushleft
 \includegraphics[width=\columnwidth]{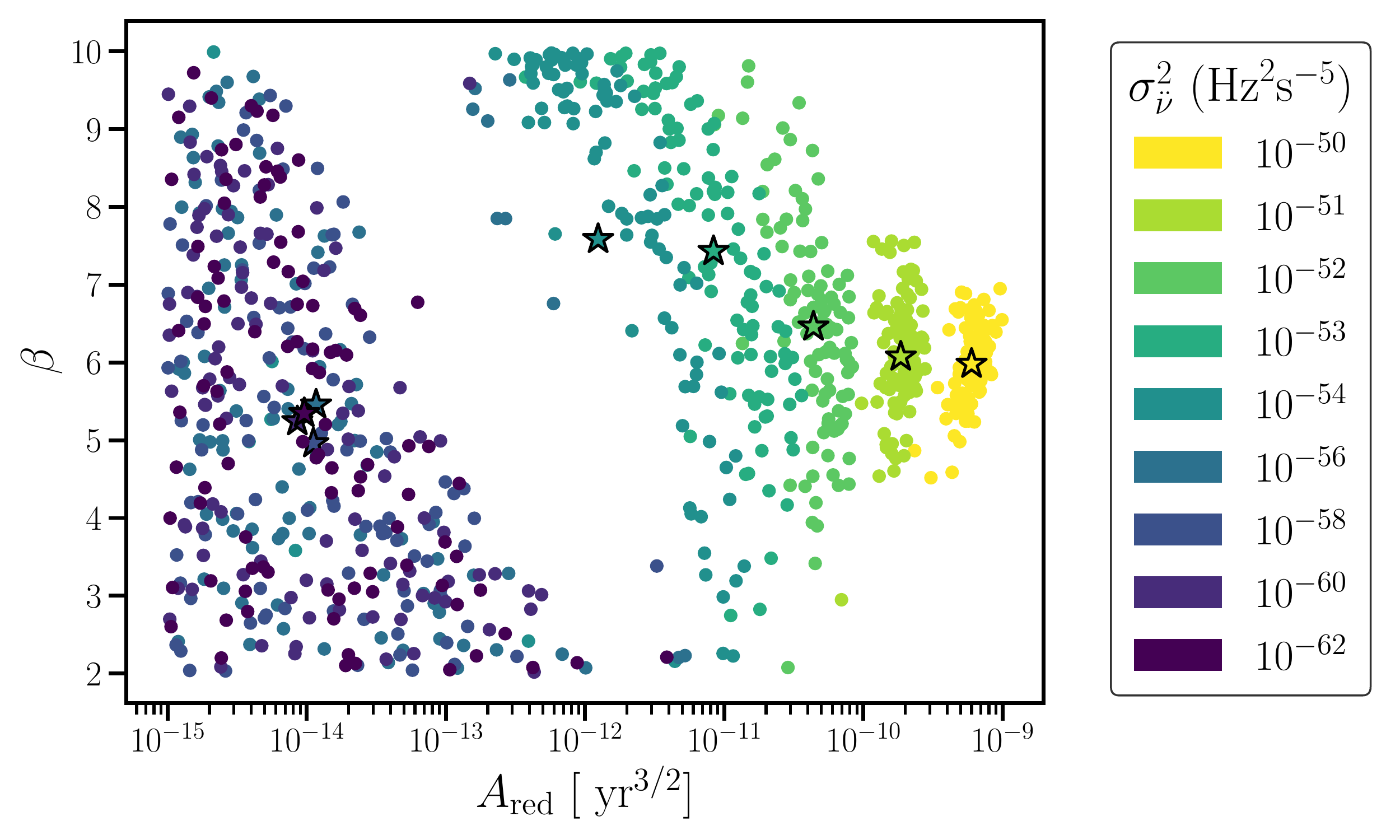}
 \caption{Relation between the injected Langevin noise amplitude squared, $\sigma_{\ddot{\nu}}^{2}$ in (\ref{Eq:Sigma_Matrix}) (units: ${\rm Hz}^{2}{\rm s}^{-5}$; see color scale) and the phase residual PSD parameters $A_{\rm red}$ (units: ${\rm yr^{3/2}}$) and $\beta$ in (\ref{Eq:Temponest_TimingNoise}) inferred by~\temponest. Each round dot corresponds to one of the $100$ random realizations analyzed in Figure~\ref{fig_subsecIV:DISP_vs_sigmas}. Stars denote $\langle A_{\rm red} \rangle$ and $\langle \beta \rangle$ for the ensemble of 100 realizations at each of the nine values $\sigma_{\ddot{\nu}}^{2} / (1 \, {\rm Hz}^2{\rm s}^{-5}) = 10^{-50}, 10^{-51}, 10^{-52},10^{-53},10^{-54},10^{-56},10^{-58},10^{-60}$, and $10^{-62}$ (indicated by the color scheme in the legend).}
\label{fig_subsecIV:Recovered_Ramp_Rslope}
\end{figure}

Figure~\ref{fig_subsecIV:Recovered_Ramp_Rslope} serves as a bridge between $\sigma^{2}_{\ddot{\nu}}$, $A_{\rm red}$ and $\beta$. For each of the 100 noise realizations analyzed in Section~\ref{subsec_IV:DISP_n_vs_sigma2}, we plot a round dot in Figure~\ref{fig_subsecIV:Recovered_Ramp_Rslope}, whose position in the plane of the plot indicates the ordered pair $(A_{\rm red},\beta)$ inferred by~\temponest, and whose color indicates $\sigma_{\ddot{\nu}}^{2}$ injected in the range $10^{-62} \leq \sigma_{\ddot{\nu}}^2 / (1\, {\rm Hz}^2{\rm s}^{-5}) \leq 10^{-50}$. For each of the nine ensembles at fixed $\sigma_{\ddot{\nu}}^{2} / (1\, {\rm Hz}^2{\rm s}^{-5}) = 10^{-50}, 10^{-51}, 10^{-52},10^{-53},10^{-54},10^{-56},10^{-58},10^{-60}$, and $10^{-62}$ (containing 100 random realizations each), we also plot the ensemble averages $\langle A_{\rm red} \rangle$ and $\langle \beta\rangle$ as open stars, with the enclosed colored dot indicating $\sigma_{\ddot{\nu}}^{2}$ via the color bar. For high timing noise $(\sigma_{\ddot{\nu}}^2 \gtrsim 10^{-51} \, {\rm Hz}^2{\rm s}^{-5})$, both $A_{\rm red}$ and $\beta$ are clustered, spanning $\sim 4.2~{\rm yr}^{3/2}$ in $A_{\rm red}$ and $4.5 \lesssim \beta \lesssim 6.9$. The clustering weakens, as $\sigma_{\ddot{\nu}}$ decreases. For example,
for $\sigma_{\ddot{\nu}}^2 = 10^{-56} \, {\rm Hz}^2{\rm s}^{-5}$, $A_{\rm red}$ and $\beta$ span $\sim 15~{\rm yr}^{3/2}$ and $2.1 \lesssim \beta \lesssim 9.7$.  It means that random variation between noise realizations produces considerable dispersion in $A_{\rm red}$ and $\beta$, even when the injected Langevin parameters $\gamma_{\ddot{\nu}}$ and $\sigma_{\ddot{\nu}}^{2}$ are the same. In other words, there is a limit to how accurately one can tie~\temponest~measurements of $A_{\rm red}$ and $\beta$ to the dynamical parameters of an underlying noise process, whether the process takes the form (\ref{Eq_SecII:model})--(\ref{Eq_subsecII:memory_less}) or something else. The stars in Figure~\ref{fig_subsecIV:Recovered_Ramp_Rslope} exhibit a tighter trend, but they are not useful in practical astronomical applications, because one observes a particular noise realization (the actual one) in any given pulsar, with no way of knowing where that realization lies compared to the ensemble average (see Section~\ref{sec:Recovered_n_stats}).

The diagonal white band in Figure~\ref{fig_subsecIV:Recovered_Ramp_Rslope}, roughly described by the line $\beta\approx-3.5 \log(A_{\rm red}/{\rm yr}^{3/2})-39$, coincides with the demarcation point between region (i) and region (ii), for a PSR J0942$-$5552-like object. We stress that Figure~\ref{fig_subsecIV:Recovered_Ramp_Rslope} is representative of PSR J0942$-$5552-like values and should be reverified with different $\nu(t_0)$ and $\dot{\nu}(t_0)$ when the time comes to apply the controlled synthetic experiments in this paper to a set of real pulsars (see also Section~\ref{subsec_IV:DISP_n_vs_f_fdot}).

\subsection{${\rm ERR}(n)$ versus $\sigma^{2}_{\ddot{\nu}}$}
\label{subsec_IV:ERRn_vs_sigma2}

As explained in Section~\ref{sec:Recovered_n_stats}, it is impossible to construct the blue and cyan histograms, presented in Figure~\ref{fig_subsecIII:hists_vs_sigmas}, for a real pulsar as only one unique noise realization is measured. Knowing ${\rm ERR}(n)$ theoretically, therefore, does not help when seeking to predict how far the measured $n$ is displaced from $n_{\rm pl}$. Instead, ${\rm DISP}(n)$ is a better proxy. For completeness, we note briefly that ${\rm ERR}(n)$ increases with $\sigma_{\ddot{\nu}}$. In region (i) from Figure~\ref{fig_subsecIV:DISP_vs_sigmas}, we find $ | {\rm ERR}(n) | \leq 0.1$. In region (ii) we find $ | {\rm ERR}(n) | \approx 0.8~(\sigma_{\ddot{\nu}}^2 / 10^{-56} \,{\rm Hz}^{2}{\rm s}^{-5})$.

\subsection{${\rm DISP}(n)$ versus $\nu(t_0)$ and $\dot{\nu}(t_0)$}
\label{subsec_IV:DISP_n_vs_f_fdot}

A natural question is whether or not the results in Sections~\ref{sec:Recovered_n_stats} and \ref{subsec_IV:DISP_n_vs_sigma2}--\ref{subsec_IV:ERRn_vs_sigma2} depend on $\nu(t_0)$ and $\dot{\nu}(t_0)$. In other words, is it possible to find a universal condition on the noise amplitude that yields $|n| \gg n_{\rm pl}$ across a large subset of the radio pulsar population, e.g.\ for a large percentage of the objects in the ATNF Pulsar Database \citep{ManchesterHobbs2005,ParthasarathyShannon2019,LowerBailes2020,ParthasarathyJohnston2020}?  Equation (\ref{Eq_SecIV:DISP_n_vs_sigma_gammaT}) suggests that the answer is yes, because ${\rm DISP}(n)$ --- and hence the condition ${\rm DISP}(n) \gg 1$ --- is independent of $\nu(t_0)$ and $\dot{\nu}(t_0)$. We check whether this is true empirically in this section by running tests on synthetic data.

To conduct the tests, we perform the following numerical experiment. We draw a random $\nu(t_{0})$ from a uniform distribution, in the arbitrary but representative range $0.3 \leq \nu(t_0) / (1\, {\rm Hz}) \leq 7.5$. Holding $\ddot{\nu}(t_{0})$ and $n_{\rm pl}$ fixed (i.e. using the injected values found in Table~\ref{Table_subsecII:example_injected_values}) we calculate $\dot{\nu}(t_{0})$ via (\ref{Eq:Intro_n}). We create a synthetic data realization, at a given $\sigma^{2}_{\ddot{\nu}}$, for the random $\nu(t_{0})$ and $\dot{\nu}(t_{0})$ pair and feed it to~\temponest~as detailed in Section~\ref{subsecII:gen_data_get_n_example}. This process is repeated $100$ times per $\sigma^{2}_{\ddot{\nu}}$ value, with the latter variable covering $10^{-56}\leq \sigma^{2}_{\ddot{\nu}} / (1 \, {\rm Hz}^{2}{\rm s}^{-5}) \leq 10^{-50}$ in steps of two decades. We calculate ${\rm DISP}(n)$ using (\ref{Eq:Dispersion_n}). Hereafter we use the subscript ``${\rm v}$'', viz. ${\rm DISP}_{\rm v}(n)$, to label the previous procedure, i.e. varying $\nu(t_{0})$ and $\dot{\nu}(t_{0})$, as opposed to ${\rm DISP}(n)$ for the unique values of $\nu(t_{0})$ and $\dot{\nu}(t_{0})$ listed in Table~\ref{Table_subsecII:example_injected_values} and presented in Section~\ref{subsec_IV:DISP_n_vs_sigma2} .

The results of the experiment are summarized in Table~\ref{Table:rand_DISP_vs_static_DISP}. ${\rm DISP}_{\rm v}(n)$ (central column in Table~\ref{Table:rand_DISP_vs_static_DISP}) is presented alongside the fraction ${\rm DISP}_{\rm v}(n)/{\rm DISP}(n)$ (rightmost column in Table~\ref{Table:rand_DISP_vs_static_DISP}). The fraction ${\rm DISP}_{\rm v}(n)/{\rm DISP}(n)$ varies from $0.8$ to $1.8$ across six decades of $\sigma^{2}_{\ddot{\nu}}$ values. In other words, ${\rm DISP}(n)$ depends weakly on $\nu(t_0)$ and $\dot{\nu}(t_0)$ in the regime of astrophysical interest. Furthermore, ${\rm DISP}_{\rm v}(n)$ approximates ${\rm DISP}(n)$ more closely, as $\sigma_{\ddot{\nu}}$ increases, and the measured $n$ grows more anomalous. That is, we obtain ${\rm DISP}_{\rm v}(n) \approx {\rm DISP}(n)$ independent of $\nu(t_0)$ and $\dot{\nu}(t_0)$ in the regime where it matters most. It is not surprising that ${\rm DISP}_{\rm v}(n)/{\rm DISP}(n)$ deviates weakly from unity for lower $\sigma^{2}_{\ddot{\nu}}$; the latter regime corresponds to region (i) in Figure~\ref{fig_subsecIV:DISP_vs_sigmas}, which does not obey the trend in (\ref{Eq_SecIV:DISP_n_vs_sigma_gammaT}). Again, however, one obtains $n\approx n_{\rm pl}$ in region (i), so the braking index is not anomalous, and ${\rm DISP}(n)$ is less consequential.  

One drawback of (\ref{Eq_SecIV:DISP_n_vs_sigma_gammaT}) when applied to real astronomical data is that the right-hand side involves $\ddot{\nu}_{\rm em}(t_0)$, which is difficult to measure accurately when timing noise masks the secular spin down --- the central point, indeed, of this paper. Of course, one can obtain a fair estimate by assuming $\ddot{\nu}_{\rm em}(t_0) = n_{\rm pl} \dot{\nu}(t_0)^2/ \nu(t_0) \sim \dot{\nu}(t_0)^2/ \nu(t_0)$ for most plausible physical torques (see Section~\ref{Sec:Introduction}), as suggested in Section~\ref{subsec_IV:DISP_n_vs_sigma2}. There is a second valid approach, however. Instead of the fractional variance ${\rm DISP}(n)$, one can consider the absolute variance 

\begin{equation}
\langle n^2 \rangle - n_{\rm pl}^2 = \frac{\sigma_{\ddot{\nu}}^2 \nu_{\rm em}(t_0)^2 } { \gamma_{\ddot{\nu}}^2 \dot{\nu}_{\rm em}(t_0)^4 T_{\rm obs} }.
\label{Eq_SecIV:Variance_sigma_gammaT}
\end{equation}

Equation (\ref{Eq_SecIV:Variance_sigma_gammaT}) does introduce new dependences on $\nu_{\rm em}(t_0)$ and $\dot{\nu}_{\rm em}(t_0)$, but the dependences are straightforward, and $\nu_{\rm em}(t_0)$ and $\dot{\nu}_{\rm em}(t_0)$ are always measured accurately in any pulsar for astrophysically plausible $\sigma_{\ddot{\nu}}$ values, unlike $\ddot{\nu}_{\rm em}(t_0)$. The reader is encouraged to select between (\ref{Eq_SecIV:DISP_n_vs_sigma_gammaT}) or (\ref{Eq_SecIV:Variance_sigma_gammaT}) to suit the application at hand.

\begin{table}
\caption{Results of varying $\nu(t_{0})$ and $\dot{\nu}(t_{0})$ on the dispersion of $n$ measurements. From left to right, the columns list the injected $\sigma^{2}_{\ddot{\nu}}$, the dispersion when varying $\nu(t_{0})$ and $\dot{\nu}(t_{0})$, denoted ${\rm DISP}_{\rm v}(n)$, and the fraction ${\rm DISP}_{\rm v}(n)/{\rm DISP}(n)$, with ${\rm DISP}(n)$ copied from Figure~\ref{fig_subsecIV:DISP_vs_sigmas}. }
\flushleft
\begin{tabular}{p{2cm}p{2.5cm}p{2.5cm}}
\hline
$\sigma^{2}_{\ddot{\nu}}/~{\rm Hz}^{2}{\rm s}^{-5}$ & ${\rm DISP}_{\rm v}(n)$ & ${\rm DISP}_{\rm v}(n)/{\rm DISP}(n)$ \\
\hline
$10^{-50}$ & $3.6\times10^{6}$ & $1.0$ \\
$10^{-52}$ & $3.0\times10^{4}$ & $0.8$ \\
$10^{-54}$ & $4.9\times10^{2}$ & $1.4$ \\
$10^{-56}$ & $9.4\times10^{0}$ & $1.8$\\
\hline
\end{tabular}
\label{Table:rand_DISP_vs_static_DISP}
\end{table}

\section{Conclusions}
\label{sec:Conclusions}

Stochastic spin wandering (i.e.\ the achromatic component of timing noise unrelated to magnetospheric and interstellar propagation) introduces random dispersion into measurements of the braking index of a rotation-powered pulsar, masking the underlying, secular (e.g.\ electromagnetic) braking torque. We quantify the masking phenomenon in this paper through a combination of analytic theory and controlled, systematic, numerical experiments based on synthetic data and the pulsar timing software~\temponest. The synthetic data are generated for the representative model $\ddot{\nu}(t)= \ddot{\nu}_{\rm em}(t) + \zeta(t)$, where $\nu_{\rm em}(t)$ satisfies $\dot{\nu}_{\rm em}(t) = K \nu_{\rm em}(t)^{n_{\rm pl}}$ with $n_{\rm pl}=3$ and $K = {\rm constant}$ in this paper for the sake of definiteness, and $\zeta(t)$ is a fluctuating, zero-mean, Langevin driver. Alternative models (e.g.\ with $K$ varying) are equally valid and can be studied within the same framework. The synthetic TOAs are fed into~\tempoDOS~and~\temponest~to produce two traditional timing solutions as a cross-check.~\temponest~output is converted into a measurement of $n$ for each random realization of the noise and compared with the injected value of $n_{\rm pl}$.

The probability distribution of the synthetic $n$ measurements, presented in Section~\ref{sec:Recovered_n_stats} and Figure~\ref{fig_subsecIII:hists_vs_sigmas}, reveals two important properties. First, the dispersion of $n$ [as measured by ${\rm DISP}(n)$, for example] for an ensemble of random noise realizations with fixed amplitude $\sigma_{\ddot{\nu}}$ is typically greater than the formal uncertainty $\Delta n$ returned by~\temponest~for a single random noise realization with the same $\sigma_{\ddot{\nu}}$ in the astrophysically relevant regime $\sigma_{\ddot{\nu}}^2 \gtrsim 10^{-56} \, {\rm Hz}^2{\rm s}^{-5}$. That is, the blue and cyan histograms in Figure~\ref{fig_subsecIII:hists_vs_sigmas} are wider than the orange histograms. Second, the dispersion of $n$ is typically greater than $n_{\rm pl}$ in the same regime. That is, $n_{\rm pl}$ falls outside the range $[n-\Delta n, n+\Delta n$] for more realizations, as $\sigma_{\ddot{\nu}}^2 \gtrsim 10^{-56} \, {\rm Hz}^2{\rm s}^{-5}$ increases. For example, at least $90\%$ of the $[n-\Delta n, n+\Delta n$] intervals include $n_{\rm pl}$ for $\sigma_{\ddot{\nu}}^2 \leq 10^{-58} \, {\rm Hz}^2{\rm s}^{-5}$. Yet, this percentage drops to $50\%$ for $\sigma^{2}_{\ddot{\nu}}=10^{-56}~{\rm Hz}^{2}{\rm s}^{-5}$ and $8\%$ for $\sigma^{2}_{\ddot{\nu}}=10^{-50}~{\rm Hz}^{2}{\rm s}^{-5}$. As an astronomical observation of a real pulsar involves analyzing a single noise realization (the real one), and there is no way to know where that realization lies within the ensemble, the practical uncertainty in measuring $n_{\rm pl}$ (as opposed to $n$) for a real pulsar is governed by ${\rm DISP}(n) \gg \Delta n$ rather than $\Delta n$.

The scaling of ${\rm DISP}(n)$ versus $\sigma_{\ddot{\nu}}$ in (\ref{Eq_SecIV:DISP_n_vs_sigma_gammaT}), derived analytically in Appendix~\ref{Appendix:Theory_anom_n} and confirmed by Monte Carlo simulations with synthetic data, is the central result of the paper. It applies in the high-noise regime $10^{-58} \leq \sigma_{\ddot{\nu}}^2 / (1 \,{\rm Hz}^{2}{\rm s}^{-5}) \leq 10^{-50}$, where anomalous braking indices $|n| \gg n_{\rm pl}$ are measured. In the low-noise regime $\sigma_{\ddot{\nu}}^2 \leq 10^{-58} \, {\rm Hz^2 \, s^{-5}}$, one obtains ${\rm DISP}(n) \lesssim 1$, and measurements return $n \approx n_{\rm pl}$ instead. Reexpressing (\ref{Eq_SecIV:DISP_n_vs_sigma_gammaT}) equivalently as (\ref{Eq_SecIV:Variance_sigma_gammaT}), one arrives at the condition

\begin{align}
    \sigma_{\ddot{\nu}}^2 \gg& 10^{-60} (\gamma_{\ddot{\nu}}/10^{-6} \, {\rm s^{-1}})^2 (\dot{\nu} / 10^{-14} \, {\rm Hz \, s^{-1}})^4(\nu / 1 \, {\rm Hz})^{-2} \nonumber\\
    &\times(T_{\rm obs} / 10^8 \, {\rm s}) \, {\rm Hz}^2{\rm s}^{-5 }
    \label{Eq_Conc:conditionSigma2ddotnu}
\end{align}

for the measured braking index $|n| \gg n_{\rm pl}$ to be anomalous, such that $n_{\rm pl}$ is likely to fall outside the measured range $[n-\Delta n, n+\Delta n]$. Equation~(\ref{Eq_Conc:conditionSigma2ddotnu}) may prove helpful in analyzing and interpreting real data in the future, because it is composed of observables: $\sigma_{\ddot{\nu}}^{2}$ can be related to the PSD of the red noise phase residuals inferred by~\temponest, as discussed in Section~\ref{subsec_IV:DISP_n_vs_TN_PSD} and Figure~\ref{fig_subsecIV:Recovered_Ramp_Rslope}, and $\gamma_{\ddot{\nu}}$ can be estimated approximately for the pulsar population as a whole from glitch recoveries or auto-correlation statistics, as discussed in Section~\ref{subsec_IV:DISP_n_vs_sigma2}. 

One may wonder whether the dispersion in $n$ measurements studied in this paper arises from the particular Brownian form of the timing noise dynamics postulated in Section~\ref{subsecII:Noisy_ts_gen}, encapsulated by equations (\ref{Eq_SecII:model})--(\ref{Eq_subsecII:memory_less}). The answer is no. The exact amount of dispersion does depend on the particular noise dynamics, as does the exact analytic form of the testable formula for ${\rm DISP}(n)$ given by (\ref{Eq_SecIV:DISP_n_vs_sigma_gammaT}) or equivalently (\ref{Eq_SecIV:Variance_sigma_gammaT}). However, the fact that the dispersion exists at all, and that the resulting braking indices are anomalous with $\vert n \vert \gg n_{\rm pl}$, are generic features of any reasonable noise process of sufficient amplitude, not just the one in Section~\ref{subsecII:Noisy_ts_gen}. We verify this by repeating the numerical experiments in Sections \ref{sec:BI_recovery_scheme}--\ref{sec:Disp_n_vs_noise} for synthetic data generated according to the default noise process offered in~\tempoDOS, whose PSD is a power law defined by an amplitude and an exponent according to (\ref{Eq:Temponest_TimingNoise}). The results are presented in Appendix~\ref{AppendixB:powerlawDISP}. We find that they are broadly the same as in Sections \ref{sec:BI_recovery_scheme}--\ref{sec:Disp_n_vs_noise}, e.g.\ the blue and cyan histograms in Figure~\ref{fig_subsecIII:hists_vs_sigmas} are unchanged qualitatively, and one still measures $\vert n \vert \gg n_{\rm pl}$ under a range of circumstances. ${\rm DISP}(n)$ for the power-law PSD in~\tempoDOS~is $\sim 20$ times smaller than predicted by (\ref{Eq_SecIV:DISP_n_vs_sigma_gammaT}) or (\ref{Eq_SecIV:Variance_sigma_gammaT}).

Ultimately the ideas in this paper should be tested with real data. This brings some challenges. The tests in Sections~\ref{sec:Recovered_n_stats} and \ref{sec:Disp_n_vs_noise}, which examine under what conditions \temponest~recovers the ``true'', underlying, secular $n_{\rm pl}$, can be performed with synthetic data only, where the injected $n_{\rm pl}$ is known. In real data, $n_{\rm pl}$ is never known independently; no quantitative, predictive, electrodynamic theory of pulsar braking exists at present \citep{MelroseYuen2016}. Nevertheless it may be possible, with sufficient data, to perform Bayesian model selection on a range of torque models in an effort to determine which one is preferred statistically. To do this, it is advantageous to convert the torque model (including stochastic spin wandering) into its state-space representation, so that a direct comparison can be made between the observed time series (e.g.\ TOAs) and the predictions of the model. Some promising steps have been made recently in this direction, for example when searching for pulsar glitches with a hidden Markov model \citep{MelatosDunn2020,DunnMelatos2022}, or when estimating the parameters of the two-component, crust-superfluid model of a neutron star by coupling a Kalman filter to expectation-maximization or Markov chain Monte Carlo algorithms \citep{MeyersO'Neill2021,MeyersMelatos2021}. It is a priority to apply the latter techniques, and others under development, to the high-quality data sets being generated by the latest generation of high-cadence pulsar timing campaigns \citep{NamkhamJaroenjittichai2019a,NamkhamJaroenjittichai2019b,LowerBailes2020,ParthasarathyBailes2021,JohnstonSobey2021}.

%%%%%%%%%%%%%%%%%%%%%%%%%%%%%%%%%%%%%%%%%%%%%%%%%%%%%%%%

\section*{Acknowledgements}

The authors thank Patrick Meyers, for discussing the Brownian model and making the {\tt baboo} package freely available, and  Liam Dunn, for guidance in the use of the~\temponest~and~\tempoDOS~software infrastructure and advice that allowed the Brownian model to qualitatively emulate the phase residuals of PSR J0942$-$5552 in Figure~\ref{fig_subsecII:example_f2_walk}. Additionally, we thank Patrick Meyers, Liam Dunn, Nicholas O'Neill, and Joe O'Leary for useful discussions regarding Section~\ref{sec:Disp_n_vs_noise}, which led ultimately to an understanding of local and nonlocal $n$ measurements. We thank the anonymous referee for helpful suggestions which improved the manuscript. This research
was supported by the Australian Research Council Centre of Excellence for Gravitational Wave Discovery (OzGrav), grant number CE170100004.  A. F. Vargas is supported by a Melbourne Research Scholarship and by the N. D. Goldsworthy Scholarship for Physics. The numerical calculations were performed on the OzSTAR supercomputer facility at Swinburne University of Technology. The OzSTAR program receives funding in part from the Astronomy National Collaborative Research Infrastructure Strategy (NCRIS) allocation provided by the Australian Government.

%%%%%%%%%%%%%%%%%%%%%%%%%%%%%%%%%%%%%%%%%%%%%%%%%%
\section*{Data Availability}

The timing solution for PSR J0942$-$5552 comes from \cite{LowerBailes2020}. All the synthetic data are generated using the open access software package {\tt baboo} available at \url{http://www.github.com/meyers-academic/baboo}~\citep{MeyersO'Neill2021}. We use~\tempoDOS~\citep{HobbsEdwards2006} and~\temponest~\citep{LentatiAlexander2014} to obtain timing solutions for the synthetic data.

%%%%%%%%%%%%%%%%%%%%%%%%%%%%%%%%%%%%%%%%%%%%%%%%%%%%%%%%%%%%%%%%%%%%%%%%%%%%%%%%%%%%%%%%%%%%%%

\twocolumn

\bibliographystyle{mnras}
\bibliography{ADSABS_bib, main_non_ads_bib}

%%%%%%%%%%%%%%%%%%%%%%%%%%%%%%%%%%%%%%%%%%%%%%%%%%

%%%%%%%%%%%%%%%%% APPENDICES %%%%%%%%%%%%%%%%%%%%%

\appendix
\section{Theory of anomalous braking indices}
\label{Appendix:Theory_anom_n}

In this appendix, we present an analytic theory of anomalous braking indices, which relates the statistics of the measured $n$ value to its secular component $n_{\rm pl}$ and to the properties of the stochastic spin wandering, including importantly the amplitude $\sigma_{\ddot{\nu}}$. The central result is a prediction for the dispersion ${\rm DISP}(n)$ as a function of $\sigma_{\ddot{\nu}}$ [see equation (\ref{Eq_SecIV:DISP_n_vs_sigma_gammaT})] and hence a condition on $\sigma_{\ddot{\nu}}$ which, when satisfied, implies $|n| \gg n_{\rm pl}$. To this end, we present in Appendix~\ref{Appendix1_subsec:Solution-2-6} the analytical solution for the Brownian model described by the system of stochastic differential equations (\ref{Eq_SecII:model})--(\ref{Eq_subsecII:memory_less}). We use the analytic solution of (\ref{Eq_SecII:model})--(\ref{Eq_subsecII:memory_less}) to calculate the statistics of $n$, when $n$ is measured in two ways: ``nonlocally'', via a finite difference formula involving $\dot{\nu}(t_0)$ and $\dot{\nu}(t_0+T_{\rm obs})$, and ``locally'', by calculating the second derivative $\ddot{\nu}$ directly from the component $X_4(t)$ of the state vector in (\ref{Eq_SecII:model}). The nonlocal and local approaches are described and justified in Appendices~\ref{Appendix1_subsec:nonLocal} and~\ref{Appendix1_subsec:Local} respectively. 

\subsection{Analytic solution of the Brownian model (\ref{Eq_SecII:model})--(\ref{Eq_subsecII:memory_less})}
\label{Appendix1_subsec:Solution-2-6}

In preparation for solving (\ref{Eq_SecII:model})--(\ref{Eq_subsecII:memory_less}), we write out the right-hand side of (\ref{Eq_SecII:model}) in the familiar Langevin form

\begin{align}
    \frac{d\phi(t)}{dt}  &= \nu(t), \label{Eq_Apndx1:diff_phi}\\
    \frac{d\nu(t)}{dt}  &= -\gamma_{\nu}[\nu(t)-\nu_{\rm em}(t)]+\dot{\nu}(t), \label{Eq_Apndx1:diff_nu}\\
    \frac{d\dot{\nu}(t)}{dt}  &= -\gamma_{\dot{\nu}}[\dot{\nu}(t)-\dot{\nu}_{\rm em}(t)]+\ddot{\nu}(t), \\
    \frac{d\ddot{\nu}(t)}{dt}  &= -\gamma_{\ddot{\nu}}[\ddot{\nu}(t)-\ddot{\nu}_{\rm em}(t)]+\dddot{\nu}_{\rm em}(t)+\xi(t), \label{Eq_Apndx1:diff_ddot_nu}
\end{align}

where $\xi(t)$ is a white noise driver, with $\langle \xi(t) \rangle =0$ and $\langle \xi(t)\xi(t') \rangle = \sigma^{2}_{\ddot{\nu}}\delta(t-t')$. In (\ref{Eq_Apndx1:diff_nu})--(\ref{Eq_Apndx1:diff_ddot_nu}), the secular braking behavior takes the form

\begin{equation}
    \nu_{\rm em}(t) = \nu_{\rm em}(t_{0})\left(1+\frac{t}{\tau}\right)^{-(n_{\rm pl}-1)^{-1}}, 
    \label{Eq_Apndx1:nu_em}
\end{equation}

where 

\begin{equation}
    \tau= -\frac{\nu_{\rm em}(t_{0})}{(n_{\rm pl}-1)\dot{\nu}_{\rm em}(t_{0})},
    \label{Eq_Apndx1:Tau}
\end{equation}

is the characteristic spin-down age, and $\dot{\nu}_{\rm em}(t),\ddot{\nu}_{\rm em}(t)$, and $\dddot{\nu}_{\rm em}(t)$ are the first, second, and third derivatives of (\ref{Eq_Apndx1:nu_em}), respectively.

Equations~(\ref{Eq_Apndx1:diff_nu})--(\ref{Eq_Apndx1:diff_ddot_nu}) are solvable by means of an integrating factor. Setting $\nu(t_{0})=\nu_{\rm em}(t_{0}), \dot{\nu}(t_{0})=\dot{\nu}_{\rm em}(t_{0})$, $\ddot{\nu}(t_{0})=\ddot{\nu}_{\rm em}(t_{0})$, and  $t_{0}=0$ without loss of generality, we obtain

\begin{align}
    \nu(t) &= \nu_{\rm em}(t)+e^{-\gamma_{\nu}t}\int_{0}^{t} dt' e^{\gamma_{\nu} t'} \left[ \dot{\nu}(t')-\dot{\nu}_{\rm em}(t')\right],\label{Eq_Apndx1:nu} \\
    \dot{\nu}(t) &= \dot{\nu}_{\rm em}(t)+e^{-\gamma_{\dot{\nu}}t}\int_{0}^{t} dt' e^{\gamma_{\dot{\nu}} t'} \left[ \ddot{\nu}(t')-\ddot{\nu}_{\rm em}(t')\right],\label{Eq_Apndx1:dot_nu} \\
    \ddot{\nu}(t) &= \ddot{\nu}_{\rm em}(t)+e^{-\gamma_{\ddot{\nu}}t} \int_{0}^{t} dt' e^{\gamma_{\ddot{\nu}}t'} \xi(t')\label{Eq_Apndx1:ddot_nu}.
\end{align}

We then integrate (\ref{Eq_Apndx1:diff_phi}) directly to obtain  

\begin{equation}
    \phi(t) = \phi(0)+\int_{0}^{t} dt'\,\nu(t').
    \label{Eq_Apndx1:phi}
\end{equation}

The initial phase is a historical accident, so we take $\phi(0)=0$ without loss of generality.

 It is straightforward to calculate the covariances of the zero-mean fluctuating variables $\delta\nu(t) = \nu(t) - \nu_{\rm em}(t)$, $\delta \dot{\nu}(t) = \dot{\nu}(t) - \dot{\nu}_{\rm em}(t)$, and $\delta\ddot{\nu}(t) = \ddot{\nu}(t) - \ddot{\nu}_{\rm em}(t)$ starting from (\ref{Eq_Apndx1:nu})--(\ref{Eq_Apndx1:ddot_nu}). For example (\ref{Eq_Apndx1:ddot_nu}) implies

 \begin{align}
     \langle \delta\ddot{\nu}(t)^2 \rangle &= e^{-2 \gamma_{\ddot{\nu}} t} \int_{0}^{t} dt' e^{\gamma_{\ddot{\nu}}t'} \int_{0}^{t} dt'' e^{\gamma_{\ddot{\nu}} t''} \langle \xi(t') \xi(t'') \rangle,  \\
     &= \frac{\sigma^{2}_{\ddot{\nu}}}{2\gamma_{\ddot{\nu}}}\left(1-e^{-2\gamma_{\ddot{\nu}}t}\right). \label{Eq_Apndx1:Var_ddot_nu}
 \end{align}
 
We combine (\ref{Eq_Apndx1:Var_ddot_nu}) with (\ref{Eq_Apndx1:dot_nu}) and (\ref{Eq_Apndx1:ddot_nu}) to obtain

\begin{align}
    \langle \delta \dot{\nu}(t)^{2}\rangle = &\frac{\sigma^{2}_{\ddot{\nu}}}{2\gamma_{\ddot{\nu}}\gamma_{\dot{\nu}}(\gamma_{\ddot{\nu}}-\gamma_{\dot{\nu}})^{2}(\gamma_{\ddot{\nu}}+\gamma_{\dot{\nu}})} \nonumber \\
    &\times \Bigl[(\gamma_{\ddot{\nu}}-\gamma_{\dot{\nu}})^{2}+4\gamma_{\ddot{\nu}}\gamma_{\dot{\nu}}e^{-(\gamma_{\ddot{\nu}}+\gamma_{\dot{\nu}})t} \nonumber \\
    & -(\gamma_{\ddot{\nu}}+\gamma_{\dot{\nu}})\left( \gamma_{\ddot{\nu}} e^{-2\gamma_{\dot{\nu}}t}+\gamma_{\dot{\nu}}e^{-2\gamma_{\ddot{\nu}}t} \right) \Bigr]
    \label{Eq_Apndx1:Var_dot_nu}
\end{align}

and
\begin{align}
    \langle \delta \dot{\nu}(t)\delta \ddot{\nu}(t) \rangle= &\frac{\sigma^{2}_{\ddot{\nu}}}{\gamma_{\ddot{\nu}}(\gamma_{\ddot{\nu}}^{2}-\gamma_{\dot{\nu}}^{2})}\Bigl[-\gamma_{\ddot{\nu}}e^{-(\gamma_{\ddot{\nu}}+\gamma_{\dot{\nu}})t} \nonumber \\
    &+\frac{\gamma_{\ddot{\nu}}}{2}\left(1+e^{-2\gamma_{\ddot{\nu}}t}\right)-\frac{\gamma_{\dot{\nu}}}{2}\left(1-e^{-2\gamma_{\ddot{\nu}}t}\right)\Bigr].
    \label{Eq_Apndx1:Covar_dot_nu_ddot_nu}
\end{align}

Equations (\ref{Eq_Apndx1:Var_ddot_nu})--(\ref{Eq_Apndx1:Covar_dot_nu_ddot_nu}) hold regardless of the form of $\nu_{\rm em}(t)$. In other words, when deriving (\ref{Eq_Apndx1:Var_ddot_nu})--(\ref{Eq_Apndx1:Covar_dot_nu_ddot_nu}), we do not need to make the approximation $t\leq T_{\rm obs} \ll \tau$.

The above analytic solution reproduces qualitatively the observed timing behavior of typical pulsars in the ATNF Pulsar Database \citep{ManchesterHobbs2005}, as illustrated in Figure~\ref{fig_subsecII:example_f2_walk} for the representative object PSR J0942$-$5552. Specifically, the analytic solution has the following properties.

\begin{enumerate}
    \item It exhibits fluctuations in $\nu(t),~\dot{\nu}(t)$ and $\ddot{\nu}(t)$ driven by $\xi(t)$ in (\ref{Eq_Apndx1:diff_ddot_nu}). The fluctuation amplitude matches the observed timing behavior of typical pulsars (e.g.\ Figure~\ref{fig_subsecII:example_f2_walk} and Table~\ref{Table_subsecII:example_injected_values}) in the following, well-defined regime: $\gamma_\nu \sim \gamma_{\dot{\nu}} \ll T_{\rm obs}^{-1} \ll \gamma_{\ddot{\nu}}$. When the foregoing conditions hold simultaneously, one obtains $| \delta\nu(t) | \ll \nu_{\rm em}(t)$, $| \delta\dot{\nu}(t) | \ll | \dot{\nu}_{\rm em}(t) |$, and  $ | \delta\ddot{\nu}(t) | \gg | \ddot{\nu}_{\rm em}(t) |$, as observed typically \citep{LowerBailes2020,ParthasarathyJohnston2020}. Incidentally, the foregoing conditions are also consistent with theoretical predictions of $\gamma_\nu$, $\gamma_{\dot{\nu}}$, and $\gamma_{\ddot{\nu}}$ based on pulsar glitch recoveries and timing noise auto-correlation studies \citep{PriceLink2012,MelatosDunn2020,MeyersMelatos2021,MeyersO'Neill2021}. In the above regime, for $t \leq T_{\rm obs}$, we approximate (\ref{Eq_Apndx1:Var_ddot_nu})--(\ref{Eq_Apndx1:Covar_dot_nu_ddot_nu}) as

\begin{align}
    \langle \delta \ddot{\nu}(t)^{2} \rangle &=  \frac{\sigma^{2}_{\ddot{\nu}}}{2\gamma_{\ddot{\nu}}} \label{Eq_Apndx1:var_ddot_nu_inf}\\
    \langle \delta \dot{\nu}(t)^{2} \rangle &=  \frac{\sigma^{2}_{\ddot{\nu}}}{2\gamma_{\dot{\nu}}\gamma_{\ddot{\nu}}^{2}}\left(1-e^{-2\gamma_{\dot{\nu}}t}\right), \label{Eq_Apndx1:var_dot_nu_inf}\\ \intertext{and}
    \langle \delta \dot{\nu}(t) \ddot{\nu}(t) \rangle &=  \frac{\sigma^{2}_{\ddot{\nu}}}{2\gamma_{\ddot{\nu}}^{2}} \label{Eq_Apndx1:covar_ddot_nu_dot_nu_inf}.
\end{align}
    \item The variances $\langle \delta \dot{\nu}(t)^{2} \rangle$ and $\langle \delta \ddot{\nu}(t)^{2}\rangle$ are bounded and small in the sense defined in (i), ensuring that pulsars do not reverse the sign of their spin or torque after birth. Moreover, the covariance $\langle \delta\ddot{\nu}(t) \delta\ddot{\nu}(t') \rangle$ behaves well for all $t$ and $t'$, does not diverge in the limit $t\rightarrow t'$, and stays bounded as $t = t' \leq T_{\rm obs}$ increases. This is important, because measuring $n$ involves measuring $\ddot{\nu}(t)$; that is, $\nu(t)$ must be differentiable twice in order to produce a well-behaved observable. In contrast, the third derivative $\dddot{\nu}(t) \propto \xi(t)$ is not differentiable, and the covariance $\langle \dddot{\nu}(t) \dddot{\nu}(t') \rangle \propto \delta(t - t')$ diverges for $t \rightarrow t'$, as expected for the highest-order derivatives in any Brownian model, but these behaviors do not affect the observable $n$; see also footnote 2.
    \item $\langle \dot{\nu}(t) \rangle$ and $\langle \ddot{\nu}(t) \rangle$ have sensible long-term values consistent with $n=n_{\rm pl}$. The ensemble average of (\ref{Eq_Apndx1:dot_nu}) yields $\langle \dot{\nu}(t) \rangle = \dot{\nu}_{\rm em}(t)$. Likewise the ensemble average of (\ref{Eq_Apndx1:ddot_nu}) yields $\langle \ddot{\nu}(t) \rangle = \ddot{\nu}_{\rm em}(t)$.
\end{enumerate}

\subsection{Nonlocal measurement of $n$}
\label{Appendix1_subsec:nonLocal}

In a deterministic system, where $\dot{\nu}(t)$ is a smooth function, $\ddot{\nu}(t)$ is approximated accurately by $\ddot{\nu}(t) \approx [\dot{\nu}(t+\Delta t)-\dot{\nu}(t)] / \Delta t$, provided that $\Delta t$ is small. In the stochastic system (\ref{Eq_SecII:model})--(\ref{Eq_subsecII:memory_less}), $\ddot{\nu}(t)$ fluctuates randomly on arbitrary short timescales, and the approximation $\ddot{\nu}(t) \approx [\dot{\nu}(t+\Delta t)-\dot{\nu}(t)]/ \Delta t$ breaks down even for small $\Delta t$. Fundamentally, this discrepancy is caused by $\xi(t)$ in (\ref{Eq_Apndx1:ddot_nu}) being nondifferentiable, as discussed in Appendix~\ref{Appendix1_subsec:Solution-2-6} and footnote 2. In general, although $X_{4}(t)=\ddot{\nu}(t)$ in (\ref{Eq_SecII:model}) exists instantaneously, there is no unique way of estimating $\ddot{\nu}(t)$ from measured values of $\dot{\nu}(t)$; the results for $\ddot{\nu}(t)$ and hence $n$ depends on exactly how the measurement is made. 

In this section, we formulate a nonlocal measurement of $n$, in which $\ddot{\nu}(t)$ is evaluated by finite differencing the time series $\dot{\nu}(t)$ at two distinct times $t_1$ and $t_2 > t_1$, which are not separated infinitesimally; indeed, the usual choice in practice is $t_2 = t_1 + T_{\rm obs}$. The nonlocal approach is consistent with standard approaches to measuring $n$ using~\temponest~for real pulsars \citep{LowerBailes2020,ParthasarathyJohnston2020}. Following equation (6) in \cite{JohnstonGalloway1999}, we calculate $n$ from

\begin{equation}
    n=1-\frac{\dot{\nu}(t_{1})\nu(t_{2})-\dot{\nu}(t_{2})\nu(t_{1})}{\dot{\nu}(t_{1})\dot{\nu}(t_{2})T_{\rm obs}},
    \label{Eq_Apndx2:n_diffstates}
\end{equation}

where $\nu(t_{1}),\dot{\nu}(t_{1})$, $\nu(t_{2})$, and $\dot{\nu}(t_{2})$ are measurements done at times $t_{1}$ and  $t_{2}=t_{1}+T_{\rm obs}$. Equation (\ref{Eq_Apndx2:n_diffstates}) is obtained by integrating $\dot{\nu}(t)=-K\nu(t)^{n}$ from $t=t_{1}$ to $t=t_{2}$ and eliminating $K$ via $K=-\dot{\nu}(t_1) / \nu(t_1)^n$. 

Consider the regime $\gamma_{\nu} \sim \gamma_{\dot{\nu}}\ll T^{-1}_{\rm obs} \ll \gamma_{\ddot{\nu}}$ identified in point (i) of Appendix~\ref{Appendix1_subsec:Solution-2-6}, which guarantees that $\delta \nu(t)$ and $\delta \dot{\nu}(t)$ are fluctuating terms with zero mean which obey $\vert \delta \nu(t) \vert \ll \vert \nu_{\rm em}(t) \vert$ and  $\vert \delta \dot{\nu}(t) \vert \ll \vert \dot{\nu}_{\rm em}(t) \vert$.  For example, we find $\delta\nu(t) \sim 10^{-8} | \nu_{\rm em}(t) \vert$ and $\delta \dot{\nu} \sim 10^{-3} \vert \dot{\nu}_{\rm em}(t) \vert$ for PSR J0942$-$5552 in Figure~\ref{fig_subsecII:example_f2_walk}. In this regime, we can write,

\begin{align}
    \frac{\nu(t_{2})}{\dot{\nu}(t_{2})} &= \frac{\nu_{\rm em}(t_{2})+\delta \nu(t_{2})}{\dot{\nu}_{\rm em}(t_{2})\left[1+\delta \dot{\nu}(t_{2})/\dot{\nu}_{\rm em}(t_{2})\right]},  \\
    &\approx \frac{\nu_{\rm em}(t_{2})}{\dot{\nu}_{\rm em}(t_{2})} \left[1+\frac{\delta \nu(t_{2})}{\nu_{\rm em}(t_{2})}-\frac{\delta \dot{\nu}(t_{2})}{\dot{\nu}_{\rm em}(t_{2})} \right], \label{Eq_Apnd2:frac_nu2_dnu_2}
\end{align}

and similarly for $\nu(t_{1})/\dot{\nu}(t_{1})$. Upon substituting (\ref{Eq_Apnd2:frac_nu2_dnu_2}) into (\ref{Eq_Apndx2:n_diffstates}), we obtain

\begin{align}
    n =&\,1 -\frac{1}{T_{\rm obs}} \frac{\nu_{\rm em}(t_{2})}{\dot{\nu}_{\rm em}(t_{2})} \left[1+\frac{\delta \nu(t_{2})}{\nu_{\rm em}(t_{2})}-\frac{\delta \dot{\nu}(t_{2})}{\dot{\nu}_{\rm em}(t_{2})} \right] \nonumber \\
    &+\frac{1}{T_{\rm obs}}\frac{\nu_{\rm em}(t_{1})}{\dot{\nu}_{\rm em}(t_{1})} \left[1+\frac{\delta \nu(t_{1})}{\nu_{\rm em}(t_{1})}-\frac{\delta \dot{\nu}(t_{1})}{\dot{\nu}_{\rm em}(t_{1})} \right]. \label{Eq_Apndx2:n_expanded_flucs_1_2}
\end{align}

The ensemble average of (\ref{Eq_Apndx2:n_expanded_flucs_1_2}) yields

\begin{align}
    \langle n \rangle &= 1 - \frac{1}{T_{\rm obs}}\left[\frac{\nu_{\rm em}(t_{2})}{\dot{\nu}_{\rm em}(t_{2})}-\frac{\nu_{\rm em}(t_{1})}{\dot{\nu}_{\rm em}(t_{1})}\right] \label{Eq_Apndx2:mean_n_tobe_npl}\\
    &= n_{\rm pl} \label{Eq_Apndx2:mean_n_is_npl},
\end{align}

where the last line follows by comparing~(\ref{Eq_Apndx2:mean_n_tobe_npl}) with (\ref{Eq_Apndx2:n_diffstates}) for $\dot{\nu}_{\rm em}(t)=-K\nu_{\rm em}^{n_{\rm pl}}(t)$. We note further that $\vert \delta \nu(t) / \nu_{\rm em}(t) \vert \ll \vert \delta \dot{\nu}(t) / \dot{\nu}_{\rm em}(t) \vert$ holds empirically for all pulsars observed to date \citep{LowerBailes2020,ParthasarathyJohnston2020}. Applying the latter inequality, and noting $\nu(t_{1})/\nu(t_{2})= 1+{\cal O}(T_{\rm obs}/\tau)\approx 1$, we combine (\ref{Eq_Apndx2:n_expanded_flucs_1_2}) and (\ref{Eq_Apndx2:mean_n_is_npl}) without the $\delta \dot{\nu}$ terms to obtain

\begin{equation}
    n-n_{\rm pl} \approx \frac{\nu_{\rm em}(t_{1})}{\dot{\nu}_{\rm em}(t_{1})^{2}}\left[ \frac{\delta \dot{\nu}(t_{2})-\delta \dot{\nu}(t_{1})}{T_{\rm obs}} \right].
    \label{Eq_Apndx2:n_minus_npl}
\end{equation}

In effect, equation (\ref{Eq_Apndx2:n_minus_npl}) calculates $n$ by replacing $\ddot{\nu}_{\rm em}(t)$ with its first order (Euler) finite difference approximation. However, the approximation is not ad hoc; it follows from the empirically justified limits taken in the lead-up to (\ref{Eq_Apndx2:n_minus_npl}).

Ultimately we are interested in the fractional dispersion across the ensemble of noise realizations, i.e. equation (\ref{Eq:Dispersion_n}). Therefore we need to calculate $\langle [ \delta \dot{\nu}(t_{2})-\delta \dot{\nu}(t_{1})]^{2} \rangle$. Setting $t=T_{\rm obs}$ in (\ref{Eq_Apndx1:var_dot_nu_inf}), with the time origin arbitrary (e.g.\ $t_{1}=t_{0}$), we arrive at

\begin{align} 
    {\rm DISP}(n) &= \frac{\sigma^{2}_{\ddot{\nu}}(1-e^{-2\gamma_{\dot{\nu}}T_{\rm obs}})}{2\gamma_{\ddot{\nu}}^{2}\gamma_{\dot{\nu}}T_{\rm obs}^{2}\ddot{\nu}_{\rm em}^{2}(t_{0})}. \label{Eq_Apndx2:DISP_n_exp_form}
\end{align}

In the astrophysically relevant regime $\gamma_{\dot{\nu}}T_{\rm obs} \ll 1$, (\ref{Eq_Apndx2:DISP_n_exp_form}) reduces to

\begin{equation}
    {\rm DISP}(n) = \frac{\sigma^{2}_{\ddot{\nu}}}{\gamma_{\ddot{\nu}}^{2}\ddot{\nu}^{2}_{\rm em}(t_{0})T_{\rm obs}}. 
    \label{Eq_Apndx2:DISP_n_vs_sigma_2gammaT}
\end{equation}

Equation (\ref{Eq_Apndx2:DISP_n_vs_sigma_2gammaT}) is identical to (\ref{Eq_SecIV:DISP_n_vs_sigma_gammaT}).   

\subsection{Local measurement of $n$}
\label{Appendix1_subsec:Local}

The nonlocal measurement of $\ddot{\nu}(t)$ and hence $n$ in Appendix~\ref{Appendix1_subsec:nonLocal} is not unique. Valid alternative recipes exist, as foreshadowed above, which may or may not yield the same result as (\ref{Eq_Apndx2:DISP_n_vs_sigma_2gammaT}) for ${\rm DISP}(n)$. One alternative is a local measurement, in which $n$ is obtained directly from the second-derivative component $X_{4}(t)=\ddot{\nu}(t)$ of the state vector in (\ref{Eq_SecII:model}), as opposed to the finite difference approximation presented in Appendix~\ref{Appendix1_subsec:nonLocal}.  

To calculate ${\rm DISP}(n)$ in this regime, we start from (\ref{Eq:Intro_n}) and write

\begin{align}
    n(t)^{2} &= \frac{\nu_{\rm em}^{2}(t)[\ddot{\nu}_{\rm em}(t)+\delta \ddot{\nu}(t)]^{2}}{[\dot{\nu}_{\rm em}(t)+\delta \dot{\nu}(t)]^{4}} \\
    &\approx n_{\rm pl}^{2}\left[1+\frac{\delta \ddot{\nu}(t)}{\ddot{\nu}_{\rm em}(t)}+\frac{\delta \ddot{\nu}(t)^{2}}{\ddot{\nu}_{\rm em}(t)^{2}}\right]\left[1-\frac{4\,\delta \dot{\nu}(t)}{\dot{\nu}_{\rm em}(t)}+\frac{10\,\delta \dot{\nu}(t)^{2}}{\dot{\nu}_{\rm em}(t)^{2}}\right] \label{Eq_Apndx1:perturb_n2}
\end{align}

 locally, noting that one has $ | \delta\dot{\nu}(t) | \ll |\dot{\nu}_{\rm em}(t) | $ empirically for all observed pulsars. Taking the ensemble average of (\ref{Eq_Apndx1:perturb_n2}), we obtain

\begin{equation}
    \langle n(t)^{2} \rangle = n_{\rm pl}^{2} \left[1+\frac{10\,\langle \delta \dot{\nu}(t)^{2} \rangle}{\dot{\nu}_{\rm em}(t)^{2}}-\frac{8\, \langle \delta \ddot{\nu}(t) \delta \dot{\nu}(t) \rangle}{\ddot{\nu}_{\rm em}(t)\dot{\nu}_{\rm em}(t)}+\frac{\langle \delta \ddot{\nu}(t)^{2}\rangle}{\ddot{\nu}_{\rm em}(t)^{2}}\right].
    \label{Eq_Apndx1:ln2r}
\end{equation}
 
Equations~(\ref{Eq_Apndx1:var_ddot_nu_inf})--(\ref{Eq_Apndx1:covar_ddot_nu_dot_nu_inf}) imply that the rightmost term in (\ref{Eq_Apndx1:ln2r}) is the biggest contributor in the astrophysically relevant regime $\gamma_{\nu} \sim \gamma_{\dot{\nu}}\ll T^{-1}_{\rm obs} \ll \gamma_{\ddot{\nu}}$ considered in Appendix~\ref{Appendix1_subsec:Solution-2-6}. Therefore to leading order (\ref{Eq:Dispersion_n}) reduces to

\begin{equation}
    {\rm DISP}(n) = \frac{\sigma^{2}_{\ddot{\nu}}}{2\gamma_{\ddot{\nu}}\ddot{\nu}_{\rm em}^{2}(t_{0})}. \label{Eq_Apndx1:DISP_n_sigma2}
\end{equation}

 Equation (\ref{Eq_Apndx1:DISP_n_sigma2}) equals (\ref{Eq_Apndx2:DISP_n_vs_sigma_2gammaT})  multiplied by the factor $\gamma_{\ddot{\nu}} T_{\rm obs} / 2$.  

One might ask: is (\ref{Eq_Apndx2:DISP_n_vs_sigma_2gammaT}) the ``true" ${\rm DISP}(n)$, or is it (\ref{Eq_Apndx1:DISP_n_sigma2})? The answer is both. The local and nonlocal definitions represent two different ``instruments" for measuring $n$. Both instruments measure $\langle n \rangle = n_{\rm pl}$ correctly (without bias) upon performing an ensemble average. However, such an ensemble average cannot be done in practice when analyzing astronomical data, because one observes a single noise realization (the actual one) from a pulsar, and there is no way to know where it lies within the ensemble. On the other hand, the spread of $n$ measurements with the nonlocal instrument is smaller by a factor $2(\gamma_{\ddot{\nu}}T_{\rm obs})^{-1}$ than with the local instrument. This is expected, because $\ddot{\nu}$ fluctuates randomly on arbitrarily short time-scales according to the Brownian model (\ref{Eq_SecII:model})--(\ref{Eq_subsecII:memory_less}), and a local (i.e.\ instantaneous) measurement of $\ddot{\nu}$ inherits greater dispersion from these fluctuations than a nonlocal measurement, which effectively smooths $\ddot{\nu}$ over $T_{\rm obs}$. 

\section{${\rm DISP}(\lowercase{n})$ for spin wandering generated with the power-law red-noise model in~{\sc tempo2}}
\label{AppendixB:powerlawDISP}

The Brownian model defined by (\ref{Eq_SecII:model})--(\ref{Eq_subsecII:memory_less}), known formally as an inhomogeneous Ornstein-Uhlenbeck process~\citep{Gardiner1994}, represents an idealized description of pulsar timing noise and is not unique. Other valid noise models exist, e.g.\ based on a Wiener process~\citep{Cordes1980}. It is therefore natural to ask whether or not the results in Sections~\ref{sec:Recovered_n_stats} and~\ref{sec:Disp_n_vs_noise} are general. The answer is yes: the formulas (\ref{Eq_SecIV:DISP_n_vs_sigma_gammaT}) and (\ref{Eq_SecIV:Variance_sigma_gammaT}) for ${\rm DISP}(n)$ as a function of $\sigma_{\ddot{\nu}}$ are specific to the Brownian model in Section~\ref{subsecII:Noisy_ts_gen}, but the existence of anomalous braking indices with $\langle n^2 \rangle^{1/2} \gg n_{\rm pl}$ arising from dispersion among random noise realizations is a general property of any spin wandering process of sufficient amplitude. In this appendix, we demonstrate the point by repeating the analysis in Section~\ref{sec:Recovered_n_stats} for synthetic data generated by the default red-noise PSD (\ref{Eq:Temponest_TimingNoise}) offered within~\tempoDOS. Specifically, we confirm that: (i) we obtain ${\rm DISP}(n) \gg 1$ for $6 \lesssim \beta \lesssim 8$ and $A_{\rm red} \gtrsim 10^{-10} \, {\rm yr^{3/2}}$ sufficiently large; and (ii) we obtain ${\rm DISP}(n) \gg 1$ even if we use (\ref{Eq:Temponest_TimingNoise}) both to generate the synthetic data with~\tempoDOS~and measure $n$ with~\temponest. That is, dispersion produces anomalous braking indices, whether the noise processes realized in the data and assumed in the analysis are the same or not, although ${\rm DISP}(n)$ is greater in the latter scenario of course. The analysis in this appendix builds on similar tests in the literature, e.g.\ in Section 3 of~\cite{ParthasarathyJohnston2020}.

To perform the experiment, we generate $100$ random realizations of synthetic data per value of $A_{\rm red}$ and $\beta$. We choose $A_{\rm red}$ and $\beta$ to be consistent with the range of $\sigma_{\ddot{\nu}}^2$ studied in Section~\ref{sec:Recovered_n_stats} and Figure~\ref{fig_subsecIV:Recovered_Ramp_Rslope} to enable a fair comparison. Specifically, we choose $[A_{\rm red} / (1 \, {\rm yr^{3/2}}), \beta] = (10^{-9.2}, 5.9)$, $(10^{-9.7},6.1),(10^{-10.3},6.5),(10^{-11.1},7.4)$, and $(10^{-11.9},7.6)$, which correspond to $\sigma^{2}_{\ddot{\nu}}/(1~{\rm Hz}^{2}{\rm s}^{-5})=10^{-50},10^{-51},10^{-52},10^{-53}$, and $10^{-54}$, respectively.  In other words, the power-law PSD (\ref{Eq:Temponest_TimingNoise}) is employed both when generating the noise in the synthetic data (with~\tempoDOS) and when analysing the synthetic data to measure $n$ (with~\temponest). The aforementioned values of $A_{\rm red}$ and $\beta$ are consistent with recovered timing noise parameters in the pulsar population~\citep{LowerBailes2020}. All simulations use the rotational parameters displayed in Table~\ref{Table_subsecII:example_injected_values}. We set~\temponest~priors as in Tables~\ref{Table:stochastic_params_table} and~\ref{Table:tempo2_params_table}, except for~$\ddot{\nu}$ whose prior is set as detailed in Section~\ref{subsecII:Bayesian_param_est_tnest}. 

\begin{figure}
\flushleft
 \includegraphics[width=\columnwidth]{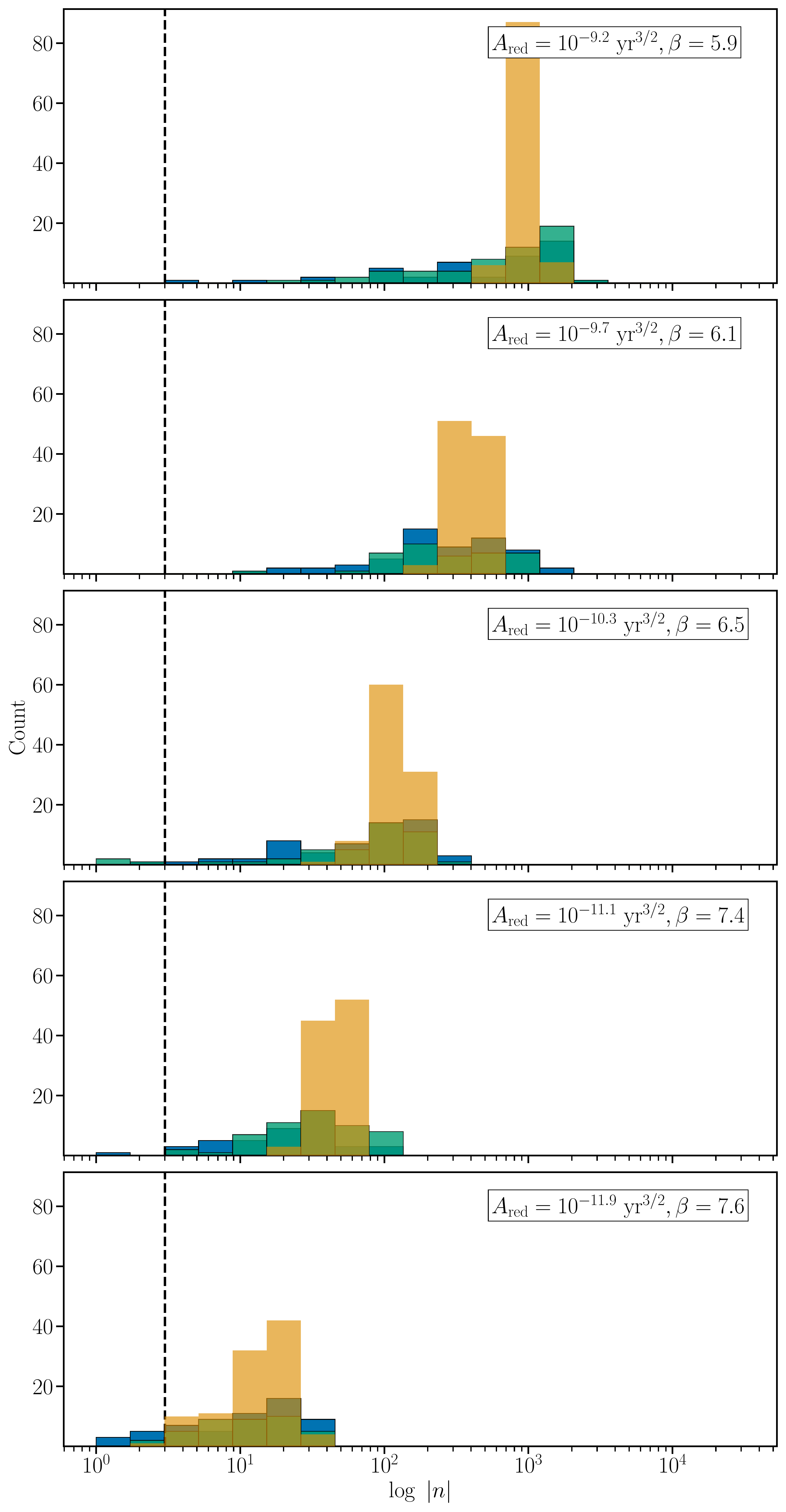}
 \caption{Validation test using the power-law noise model (\ref{Eq:Temponest_TimingNoise}) to both generate and analyze the timing data. The format is analogous to Figure~\ref{fig_subsecIII:hists_vs_sigmas}, viz.\ distributions of $n$ measurements (blue histograms for $n>0$, cyan histograms for $n<0$) and their formal uncertainties $\Delta n$ (orange histograms) reported by~\temponest~through~(\ref{Eq:Intro_n}) and~(\ref{Eq:Uncertainty_n}) for a representative sample of $A_{\rm red}$ and $\beta$ values. All panels are constructed from $10^{2}$ random realizations of synthetic data from~\tempoDOS~and the rotational parameters in Table~\ref{Table_subsecII:example_injected_values}. The black dotted line represents $n_{\rm pl}=3$. The averages of $n$ and $\Delta n$ are $2.9$ and $14.7$, and $-150$ and $982.8$, for the bottom and top panels, respectively. In the bottom and top panels, the number of measurements satisfying $n-\Delta n \leq n_{\rm pl}\leq n+\Delta n$ are $55$ and $57$, respectively. This is in contrast to $41$ and $8$ measurements for the same panels in Figure~\ref{fig_subsecIII:hists_vs_sigmas}.}
\label{fig_appendixIV:hists_LT_vs_sigmas}
\end{figure}

\noindent Figure~\ref{fig_appendixIV:hists_LT_vs_sigmas} summarizes the results of the above test in the same format as Figure~\ref{fig_subsecIII:hists_vs_sigmas}. Each panel features three histograms: a blue one, which displays the distribution of measured $n>0$ values, a cyan one, which displays the distribution of $n<0$ values, and an orange one, which displays the distribution of formal uncertainties $\Delta n$ reported by~\temponest~via (\ref{Eq:Uncertainty_n}). The panels are arranged from top to bottom in order of decreasing $A_{\rm red}$, which is analogous to decreasing $\sigma_{\ddot{\nu}}$ (effective value) in Figure~\ref{fig_subsecIII:hists_vs_sigmas}. The figure makes three key points. First, the blue and cyan histograms are wider than the orange histograms for $A_{\rm red} \gtrsim 10^{-11.9} \, {\rm yr^{3/2}}$, as in Figure~\ref{fig_subsecIII:hists_vs_sigmas}; that is, $\Delta n$ is smaller than the dispersion arising from the ensemble of random realizations, even though the same timing noise model (\ref{Eq:Temponest_TimingNoise}) is used to both generate and analyze the data. Second, anomalous braking indices with $|n| \gg n_{\rm pl}$ are measured routinely for $A_{\rm red} \gtrsim 10^{-9.2} \, {\rm yr^{3/2}}$, as in Figure~\ref{fig_subsecIII:hists_vs_sigmas}, and $n$ can take either sign, in line with the findings in Section~\ref{sec:Recovered_n_stats} and observational studies~\citep{JohnstonGalloway1999,ChukwudeChidiOdo2016,ParthasarathyJohnston2020,LowerBailes2020}. Third, ${\rm DISP}(n)$ is systematically smaller by a factor of $\sim 20$ than in Figure~\ref{fig_subsecIII:hists_vs_sigmas}, because the same timing noise model (\ref{Eq:Temponest_TimingNoise}) is used to both generate and analyze the data.

\noindent Let us quantify briefly the three points above. The FWHMs of the summed blue and cyan histograms grow with the amplitude of the power-law timing noise. The FWHM rises from $11n_{\rm pl}$ for the bottom panel to $823n_{\rm pl}$ for the top panel. The percentage of recovered $n\pm\Delta n$ intervals that include $n_{\rm pl}$ is $55\%$ and $57\%$ for the bottom and the top panels, respectively. ${\rm DISP}(n)$ also grows, from ${\rm DISP}(n)=35$ for the bottom panel to ${\rm DISP}(n)=1.3\times10^{5}$ for the top panel, as the histograms shift rightward along the logarithmic horizontal axis. The measured values of ${\rm DISP}(n)$, going from the top to the bottom panel, are $28,10,23,20$, and $10$ times smaller than those in Sections~\ref{sec:Recovered_n_stats} and~\ref{sec:Disp_n_vs_noise}  for the corresponding $\sigma_{\ddot{\nu}}^{2}$ value.

\noindent In this appendix, as in Sections~\ref{sec:Recovered_n_stats} and~\ref{sec:Disp_n_vs_noise},~\temponest~overestimates $\vert \ddot{\nu} \vert$ and underestimates $\Delta \ddot{\nu}$ on average across the ensemble of trials. Consequently, even when the same model generates and analyzes the data, the measured braking index can be anomalous, with $n_{\rm pl}$ falling outside of the measured range $[n-\Delta n, n+\Delta n]$. However, the incidence of anomalous braking indices is higher (i.e.\ ${\rm DISP}(n)$ is greater), when different models are used to generate and analyze the data, as in Sections~\ref{sec:Recovered_n_stats} and~\ref{sec:Disp_n_vs_noise}.

% Don't change these lines
\bsp	% typesetting comment
\label{lastpage}
\end{document}